\documentclass[%
 aip,
 jcp,%
 amsmath,amssymb,
reprint,%
]{revtex4-2}

\usepackage{graphicx}% Include figure files
\usepackage{dcolumn}% Align table columns on decimal point
\usepackage{bm}% bold math
\usepackage{bbm}
\usepackage{hyperref}
\usepackage{comment}
\usepackage{color}
\def\rrangle{\rangle\rangle}
\def\llangle{\langle\langle}

\begin{document}

\title{Self-consistent approach to the dynamics of excitation energy transfer in multichromophoric systems}% Force line breaks with \\

\author{Veljko Jankovi\'c}
 \affiliation{Institute of Physics Belgrade, University of Belgrade, Pregrevica 118, 11080 Belgrade, Serbia}
 \thanks{Corresponding author}
 \email{veljko.jankovic@ipb.ac.rs}
\author{Tom\'a\v s Man\v cal}
 \affiliation{Faculty of Mathematics and Physics, Charles University, Ke Karlovu 5, 121 16 Prague 2, Czech Republic}

\begin{abstract}
% version in which abbreviations are abandoned
Computationally tractable and reliable, albeit approximate, methods for studying exciton transport in molecular aggregates immersed in structured bosonic environments have been actively developed.
Going beyond the lowest-order (Born) approximation for the memory kernel of the quantum master equation typically results in complicated and possibly divergent expressions.
Starting from the memory kernel in the Born approximation, and recognizing the quantum master equation as the Dyson equation of Green's functions theory, we formulate the self-consistent Born approximation to resum the memory-kernel perturbation series in powers of the exciton--environment interaction.
Our formulation is in the Liouville space and frequency domain and handles arbitrary exciton--environment spectral densities.
In a molecular dimer coupled to an overdamped oscillator environment, we conclude that the self-consistent cycle significantly improves the Born-approximation energy-transfer dynamics.
The dynamics in the self-consistent Born approximation agree well with the solutions of hierarchical equations of motion over a wide range of parameters, including the most challenging regimes of strong exciton--environment interactions, slow environments, and low temperatures.
This is rationalized by the analytical considerations of coherence-dephasing dynamics in the pure-dephasing model.
We find that the self-consistent Born approximation is good (poor) at describing energy transfer modulated by an underdamped vibration resonant (off-resonant) with the exciton energy gap.
Nevertheless, it reasonably describes exciton dynamics in the seven-site model of the Fenna--Matthews--Olson complex in a realistic environment comprising both an overdamped continuum and underdamped vibrations.
\end{abstract}

\maketitle

\section{Introduction}
The light absorption and thus initiated excitation energy transfer (EET) in molecular aggregates constitute the first steps of solar-energy conversion in both natural~\cite{RevModPhys.90.035003,SciAdv.6.eaaz4888} and artificial~\cite{AdvEnergyMater.7.1700236,ChemRev.116.13279,SciAdv.7.eabh4232,JChemPhys.156.185102} systems.
The EET takes place in a complex dynamic spatiotemporal landscape stemming from the competition of interactions promoting exciton delocalization (resonance coupling between molecules) and localization (static and dynamic disorder).~\cite{CurrOpinChemBiol.47.1}
As the energy scales of these counteracting interactions are typically comparable to one another,~\cite{AnnuRevCondensMatterPhys.3.333,AnnuRevPhysChem.66.69} theoretical descriptions of EET dynamics are quite challenging.

When approached from the perspective of the theory of open quantum systems,~\cite{Breuer-Petruccione-book,Weiss-book} the challenge transforms into describing non-Markovian quantum dynamics of excitons interacting with their environment.
As standard theories (such as Redfield~\cite{REDFIELD19651} and F\"{o}rster~\cite{Foerster1964} theories) do not meet this challenge,~\cite{JChemPhys.130.234110} various numerically exact methods have been developed.
Two most common foundations of these are (i) the Feynman--Vernon influence functional theory~\cite{AnnPhys.24.118} and (ii) the Nakajima--Zwanzig [time-convolution (TC)] quantum master equation (QME).~\cite{JChemPhys.33.1338,ProgTheorPhys.20.948,ChinJChemPhys.34.497}
The approaches rooted in (i) include the hierarchical equations of motion (HEOM)~\cite{PhysRevE.75.031107,JChemPhys.130.234111,JChemPhys.153.020901} and a host of path-integral and process-tensor-based methods.~\cite{JChemPhys.102.4600,ChemPhysLett.478.234,JChemPhys.149.214108,JChemPhys.152.041104,JChemPhys.138.214111,NatCommun.9.3322,PRXQuantum.3.010321,NatPhys.18.662}
Each of these approaches develops a different representation of the so-called exact reduced evolution superoperator (or the dynamical map) $\mathcal{U}(t)$, which becomes its central object.
Meanwhile, the main aim of the approaches originating from (ii) is to evaluate the exact memory-kernel superoperator $\mathcal{K}(t)$.~\cite{JChemPhys.119.12063,JChemPhys.125.044106,JPhysChemB.125.9834}
All the above-referenced approaches are in general computationally intensive.
Their applications to realistic EET models, which feature a larger number of chromophores and/or structured spectral densities (SDs) of the exciton--environment interaction extracted from experiments or atomistic simulations,~\cite{PhysChemChemPhys.17.25629,JPhysChemLett.7.3171,PhysChemChemPhys.20.3310,JPhysChemB.121.10026,JChemPhys.156.120901,PhotosynthRes.156.147} can thus be impractical.
A need for computationally less demanding and reliable, although approximate, approaches to study EET dynamics in realistic multichromophoric models cannot be overemphasized.

Recent years have witnessed the emergence of many such approaches, among others quantum--classical methods,~\cite{JPhysCondensMatter.27.073201,JPhysChemA.123.1186,JChemPhys.151.224109,JChemPhys.152.244109,JPhysChemB.125.5601,JChemTheoryComput.17.7157} and related quantum--chemical approaches to nonadiabatic dynamics (e.g., the surface hopping).~\cite{JChemPhys.139.014104,JChemPhys.158.104111,runeson2023exciton}
One can also devise approximations to $\mathcal{U}$ and $\mathcal{K}$ starting from the formally exact expressions of (i) and (ii), respectively.
For example, the former give rise to cumulant-expansion-based approaches,~\cite{JChemPhys.142.094106,JPhysChemB.124.8610,JChemPhys.157.095103,JChemTheoryComput.11.415} while the latter result in various approximations to the exact memory kernel.~\cite{JChemPhys.150.034101,JChemPhys.154.204109}
The authors of Ref.~\onlinecite{PhysRevE.86.011915} noted that the second-order TC (TC2) QME, which retains only the lowest-order (second-order) term $\mathcal{K}^{(2)}(t)$ in the expansion of $\mathcal{K}$ in powers of the exciton--environment interaction, can qualitatively reproduce the main features of EET dynamics in multichromophoric systems.
The most obvious improvement over the TC2 QME, also known as the (second) Born approximation (BA),~\cite{May-Kuhn-book} is to retain some of the higher-order contributions to $\mathcal{K}$. 
Such a route generally leads to involved expressions that might not always be convergent and that can be practically evaluated only for the simplest models, where they show some improvements over the original theory.~\cite{JChemPhys.94.4391,PhysRevA.41.6596,PhysRevE.55.2328,PhysRevA.59.1633,JChemPhys.116.2705,JChemPhys.139.044102,JChemPhys.148.164101,JChemPhys.151.044110}
While already the lowest-order kernel of polaron-transformed QMEs~\cite{JChemPhys.129.101104,JChemPhys.135.034105,JChemPhys.135.114501,JChemPhys.157.104107} can significantly improve over TC2 QME, decent results can also be obtained by partially resumming the perturbation expansion for $\mathcal{K}$ using only low-order terms as the input.
Partial resummations are most commonly performed in the frequency domain, and one usually applies Pad\'e or Landau--Zener resummation schemes to $\mathcal{K}^{(2)}(\omega)$ and $\mathcal{K}^{(4)}(\omega)$.~\cite{JChemPhys.141.054112,JChemPhys.148.164101}
Another possibility, the self-consistent resummation schemes strongly rooted in condensed-matter physics,~\cite{pinescontribution,puff-whitfield-contribution,Bruus-Flensberg-book,Sadovskii-book} has received even less attention in this context.

Successful self-consistent improvements over the TC2 QME have been reported in the context of quantum theory of electronic transport through molecular junctions.~\cite{PhysLettA.357.449,JPhysChemC.114.20362,JChemPhys.140.244111,Li2016}
These approaches have mainly considered purely electronic quantum transport, when the role of the bath is played by electronic leads.
Their applications to the quantum transport in the presence of both electronic leads and molecular vibrations are much more recent and scarcer.~\cite{JChemPhys.152.064103}
A proposal for a self-consistent description of the dynamics of a general open quantum system has been made only very recently.~\cite{scarlatella2023selfconsistent}
It leans on a diagrammatic representation of the perturbation series for $\mathcal{U}(t)$, which has indeed appeared in the context of open quantum dynamics, see, for example, Refs.~\onlinecite{PhysRevA.41.6676,ChemPhys.347.185,JChemPhys.153.244122}.
The novelty of the approach by Scarlatella and Schir\`o~\cite{scarlatella2023selfconsistent} lies in the subsequent formulation of the perturbation series for $\mathcal{K}(t)$, which is related to $\mathcal{U}(t)$ via the TC QME.
In the context of the theory of Green's functions in quantum many-body systems,~\cite{Bruus-Flensberg-book,Sadovskii-book,Mahanbook} the memory kernel $\mathcal{K}(t)$ is analogous to the self-energy, whereas the TC QME plays the role of the Dyson equation.
Considering the zero-temperature dynamics of the spin--boson model, Scarlatella and Schir\`o~\cite{scarlatella2023selfconsistent} find that performing the self-consistent cycle starting from the BA memory kernel $\mathcal{K}^{(2)}$ produces promising results.
However, they consider only the zero temperature and SDs commonly used when studying the spin--boson model in its narrowest sense.~\cite{RevModPhys.59.1}
Moreover, their time-domain formulation of the self-consistent Born approximation (SCBA) involves solving an integrodifferential equation in each iteration of the cycle.

This study explores the applicability of the self-consistent memory-kernel resummation scheme introduced in Ref.~\onlinecite{scarlatella2023selfconsistent} to EET dynamics in multichromophoric aggregates.
For the sake of completeness, we first reconsider the theoretical developments of Ref.~\onlinecite{scarlatella2023selfconsistent}, and make them technically simpler by working in the Liouville space and replacing the above-mentioned integrodifferential equation in the time domain by a matrix equation in the frequency domain.
In a molecular dimer coupled to an overdamped phonon environment, we find that the self-consistent cycle greatly improves the original BA (TC2 QME) over a wide range of dimer parameters.
This success of the SCBA is rationalized by considering the analytically tractable example of coherence dephasing in the pure-dephasing model.
Remarkably, we find that the SCBA remains reliable even in the generally difficult regimes of strong exciton--environment interactions and/or slow environments, when multiple environmentally assisted processes dominate the dynamics,~\cite{JChemPhys.132.194111,PhysRevA.96.032105} as well as at low temperatures.
Considering exciton dynamics modulated by a single underdamped vibrational mode, we conclude that the SCBA delivers good results only when the vibrational energy is resonant with the exciton--energy gap.
Upon including overdamped phonon continuum on top of a number of underdamped modes, we find that the SCBA delivers decent results for EET dynamics in the seven-site model of the FMO complex interacting with the realistic environment extracted from atomistic simulations.

This paper is organized as follows.
Section~\ref{Sec:theory} introduces our theoretical framework, whose applicability to EET dynamics is assessed in Sec.~\ref{Sec:results}.
Section~\ref{Sec:summary} summarizes our main findings and discusses prospects for future work.

\section{Theoretical considerations}
\label{Sec:theory}
\subsection{Model}
We model EET dynamics in an aggregate composed of $N$ chromophores using the Frenkel--Holstein Hamiltonian,
\begin{equation}
\label{Eq:def_H_general}
    H=H_S+H_B+H_{S-B}.
\end{equation}
We take into account only the ground and one excited state on each chromophore, so that the purely electronic part $H_S$ reads as
\begin{equation}
\label{Eq:def_H_S}
    H_S=\sum_{n}\varepsilon_n|n\rangle\langle n|+\sum_{m\neq n}J_{mn}|m\rangle\langle n|.
\end{equation}
Here, $|n\rangle$ is the collective singly excited state residing on chromophore $n$ (all the other chromophores are unexcited), $\varepsilon_n$ is the energy of the vertical transition from the ground state to the excited state on chromophore $n$, while $J_{mn}$ is the resonance (typically dipole--dipole) coupling between chromophores $m$ and $n$.
The aggregate is in contact with the environment modeled as collections of mutually independent harmonic oscillators associated with each chromophore,
\begin{equation}
    H_B=\sum_{n\xi}\omega_{n\xi}b_{n\xi}^\dagger b_{n\xi}.
\end{equation}
Here, $\omega_{n\xi}$ is the frequency of oscillator $\xi$ on chromophore $n$, while bosonic operators $b_{n\xi}^\dagger$ ($b_{n\xi}$) create (annihilate) the corresponding oscillation quantum and obey $[b_{n\xi},b_{m\xi'}]=0,[b_{n\xi},b_{m\xi'}^\dagger]=\delta_{nm}\delta_{\xi\xi'}$.
The exciton--environment interaction is of the Holstein type, i.e., local and linear in oscillator displacements,
\begin{equation}
\label{Eq:H_S_B}
    H_{S-B}=\sum_{n\xi}g_{n\xi}V_n\left(b_{n\xi}+b_{n\xi}^\dagger\right),
\end{equation}
where $V_n=|n\rangle\langle n|$.
Its strength, determined by the interaction constants $g_{n\xi}$, is more conveniently described using the reorganization energy on chromophore $n$,
\begin{equation}
    \lambda_n=\sum_\xi\frac{g_{n\xi}^2}{\omega_{n\xi}}=\int_{-\infty}^{+\infty}\frac{d\omega}{2\pi}\frac{\mathcal{J}_n(\omega)}{\omega},
\end{equation}
where the SD of the exciton--environment interaction
\begin{equation}
    \mathcal{J}_n(\omega)=\pi\sum_\xi g_{n\xi}^2\left[\delta(\omega-\omega_{n\xi})-\delta(\omega+\omega_{n\xi})\right]
\end{equation}
is typically a continuous function of $\omega$.
We focus on exciton dynamics starting from the factorized initial condition
\begin{equation}
\label{Eq:factorized-initial-condition}
    W(0)=\rho(0)\rho_B^\mathrm{eq},
\end{equation}
where $\rho(0)$ is the excitonic reduced density matrix (RDM) at the initial instant $t=0$, while
\begin{equation}
    \rho_B^\mathrm{eq}=\frac{e^{-\beta H_B}}{\mathrm{Tr}_\mathrm{B}\:e^{-\beta H_B}}
\end{equation}
represents the state of the environment (equilibrium at temperature $T=\beta^{-1}$) with no excitons present.
Within the Condon approximation,~\cite{Mukamel-book} this choice of the initial condition leads to the dynamics that can be probed in ultrafast nonlinear spectroscopies.~\cite{ChemPhys.532.110663}
As our formalism deals with Green's function, we can still reconstruct the dynamics under an arbitrary excitation condition as long as the light--matter interaction is weak and the Condon approximation is valid.~\cite{NewJPhys.12.065044,ChemPhys.439.100,ChemPhys.532.110663}   
The only environmental quantity influencing the reduced excitonic dynamics starting from Eq.~\eqref{Eq:factorized-initial-condition} is the displacement autocorrelation function~\cite{AnnPhys.24.118}
\begin{equation}
\label{Eq:autocorrelation}
\begin{split}
    C_n(t)&=\sum_\xi g_{n\xi}^2\mathrm{Tr}_B\left\{[b_{n\xi}(t)+b_{n\xi}^\dagger(t)](b_{n\xi}+b_{n\xi}^\dagger)\rho_B^\mathrm{eq}\right\}\\
    &=\int_{-\infty}^{+\infty}\frac{d\omega}{\pi}e^{-i\omega t}\frac{\mathcal{J}_n(\omega)}{1-e^{-\beta\omega}}.
\end{split}
\end{equation}
The time dependence in Eq.~\eqref{Eq:autocorrelation} is taken with respect to $H_B$, i.e., $b_{n\xi}(t)=e^{-i\omega_{n\xi}t}b_{n\xi}$.

\subsection{Real-time diagrammatic representation: Green's superoperator and self-energy superoperator}
The assumption embodied in Eq.~\eqref{Eq:factorized-initial-condition} permits us to define the exact reduced evolution superoperator (dynamical map) $\mathcal{U}(t)$ by
\begin{equation}
\label{Eq:def_dynamical_map}
    |\rho(t)\rrangle=\mathcal{U}(t)|\rho(0)\rrangle.
\end{equation}
Equation~\eqref{Eq:def_dynamical_map} is formulated in the Liouville space,~\cite{Mukamel-book} in which the excitonic RDM at $t>0$, represented by the operator $\rho(t)$ in the Hilbert space, becomes the $N^2$-component vector $|\rho(t)\rrangle$.
The reduced evolution superoperator is then a tetradic quantity comprising $N^4$ entries $\llangle e'_2e'_1|\mathcal{U}(t)|e_2e_1\rrangle$, where $\{|e\rangle\}$ is an arbitrary basis in the single-exciton manifold. 
$\mathcal{U}(t)$ is formally expressed as (see, for example, Ref.~\onlinecite{PhysRevB.82.235307} and references therein)
\begin{equation}
\label{Eq:U_formal}
    \mathcal{U}(t)=e^{-i\mathcal{L}_{S}t}\mathrm{Tr}_B\left\{\mathcal{T}_te^{-i\int_0^t ds\:\mathcal{L}_{S-B}^{(I)}(s)}\rho_B^\mathrm{eq}\right\},
\end{equation}
where the Liouvillian $\mathcal{L}_a$ associated with the term $H_a$ ($a=S,B,S-B$) in Eq.~\eqref{Eq:def_H_general} is defined by its action on an arbitrary Liouville-space vector $|O\rrangle$ corresponding to the Hilbert-space operator $O$, $\mathcal{L}_a|O\rrangle\leftrightarrow [H_a,O]$, while the interaction-picture counterpart of $\mathcal{L}_{S-B}$ is
\begin{equation}
    \mathcal{L}_{S-B}^{(I)}(t)=e^{i(\mathcal{L}_S+\mathcal{L}_B)t}\mathcal{L}_{S-B}e^{-i(\mathcal{L}_S+\mathcal{L}_B)t}.
\end{equation}
The time-ordering sign $\mathcal{T}_t$ imposes the chronological order (latest to the left) among the Liouvillians in the expansion of Eq.~\eqref{Eq:U_formal} in powers of the exciton--environment interaction.
The Liouville-space approach adopted here is somewhat different from standard real-time approaches to the RDM of a particle in an oscillator environment,~\cite{scarlatella2023selfconsistent,Rammer_1998} which consider the forward and backward Hilbert-space evolution operators separately.
Dealing with superoperators, we simplify the formalism, as we consider only the forward evolution in the Liouville space.~\cite{PhysRep.465.191} 
The average in Eq.~\eqref{Eq:U_formal} can be performed term-by-term using Wick's theorem,~\cite{Rammer_1998} and only even-order powers in $\mathcal{L}_{S-B}$ remain.
The order $2k$ ($k\geq 1$) consists of $(2k-1)!!$ terms.
Further manipulations usually proceed in two different manners.

The first possibility is to observe that, because of the $\mathcal{T}_t$ sign, all the terms appearing in any given order are mutually identical.~\cite{JChemPhys.153.244122}
Retaining the $\mathcal{T}_t$ sign, one obtains the well-known Feynman--Vernon expression,~\cite{JChemPhys.130.234111,JChemPhys.153.244122}
\begin{equation}
 \label{Eq:U_t_FV}
    \mathcal{U}(t)=e^{-i\mathcal{L}_St}\mathcal{T}_te^{-\Phi(t)},
\end{equation}
with
\begin{equation}
\label{Eq:def_Phi_t}
\begin{split}
    &\Phi(t)=\sum_n\int_0^t ds_2\int_0^{s_2}ds_1\:V_n^{(I)}(s_2)^\times\\
    &\left[C_n^r(s_2-s_1)V_n^{(I)}(s_1)^\times+iC_n^i(s_2-s_1)V_n^{(I)}(s_1)^\circ\right].
\end{split}
\end{equation}
In Eq.~\eqref{Eq:def_Phi_t}, the superoperators $V^\times$ and $V^\circ$ are defined by the correspondences $V^\times|O\rrangle\leftrightarrow[V,O]$ and $V^\circ|O\rrangle\leftrightarrow\{V,O\}$ (anticommutator), respectively.
Assuming that $C_n(t)$ can be decomposed into a number of exponentially decaying terms, Eqs.~\eqref{Eq:U_t_FV} and~\eqref{Eq:def_Phi_t} serve as the starting point for the HEOM method.

\begin{figure*}[htbp!]
    \centering
    \includegraphics[width=0.9\textwidth]{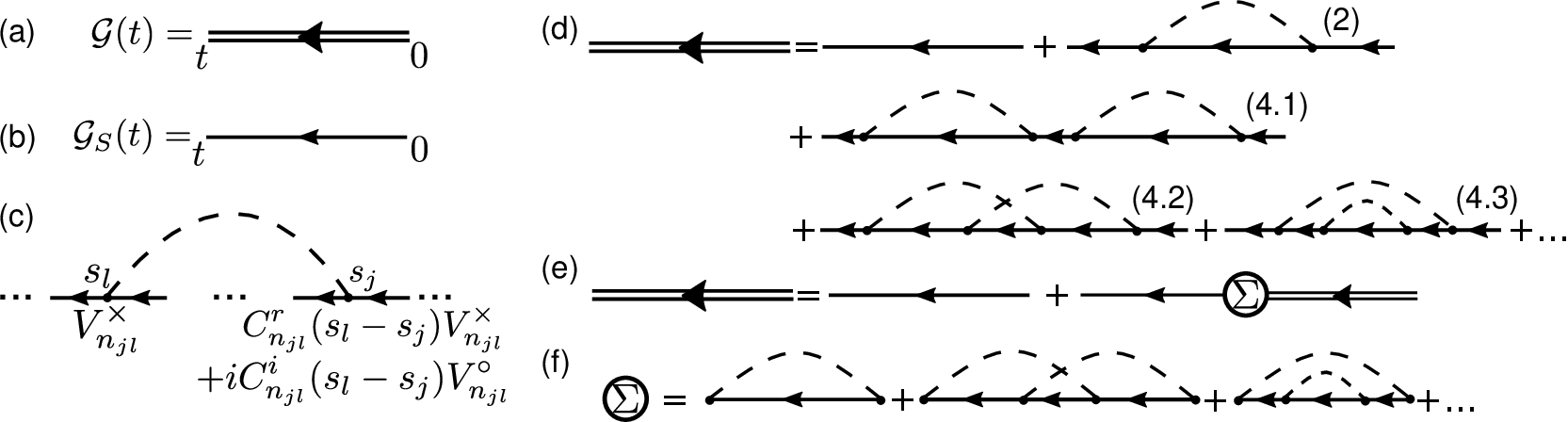}
    \caption{Diagrammatic representation of (a) the interacting Green's superoperator and (b) the free-exciton Green's superoperator.
    (c) In an arbitrary diagram, the environmental assistance starting at $s_j$, ending at $s_l$ ($t\geq s_l\geq s_j\geq 0$), and involving the environment that surrounds chromophore $n_{jl}$ is represented by a dashed circumference connecting $s_j$ and $s_l$.
    The connections of the circumference with free-exciton lines, represented by full dots, correspond to the superoperators in Eqs.~\eqref{Eq:super_s_l} and~\eqref{Eq:super_s_j}.
    Diagrammatic representation of (d) the perturbation expansion for $\mathcal{G}(t)$ up to the fourth order in the exciton--environment interaction,
    (e) the Dyson equation [Eq.~\eqref{Eq:Dyson_eq_1}], and
    (f) the self-energy superoperator up to the fourth order in the exciton--environment interaction.}
    \label{Fig:diags_final}
\end{figure*}

The possibility we opt for here is to make the time ordering explicit in each term, in which case the terms in any given order appear as different.
Further developments are facilitated by considering the (retarded) evolution superoperator [the (retarded) Green's function or Green's superoperator]~\cite{Mukamel-book}
\begin{equation}
\label{Eq:def-G-t}
    \mathcal{G}(t)=-i\theta(t)\mathcal{U}(t)
\end{equation}
instead of $\mathcal{U}(t)$.
Each term in order $2k$ of the expansion of $\mathcal{G}(t)$ can be represented by a diagram [Fig.~\ref{Fig:diags_final}(d)] comprising a total of $2k+2$ (2 terminal and $2k$ internal) instants,
$$s_{2k+1}=t\geq s_{2k}\geq\dots\geq s_1\geq 0=s_0,$$
and $k$ chromophore indices $n_k,\dots,n_1$ associated with $k$ pairs selected from the set $\{s_{2k},\dots,s_1\}$ by the application of Wick's theorem.
The consecutive instants $s_{j+1}$ and $s_j$ ($0\leq j\leq 2k+1$) of a diagram are connected by a straight line directed from $s_j$ to $s_{j+1}$ [Fig.~\ref{Fig:diags_final}(d)] and represent the (retarded) free-exciton Green's superoperator [Fig.~\ref{Fig:diags_final}(b)]
\begin{equation}
    \mathcal{G}_S(t)=-i\theta(t)e^{-i\mathcal{L}_St}.
\end{equation}
The directed double-line represents $\mathcal{G}(t)$ [Fig.~\ref{Fig:diags_final}(a)].
The dashed circumferences connect the instants $s_l$ and $s_j$ ($s_l\geq s_j$) that are paired by Wick's theorem; see Fig.~\ref{Fig:diags_final}(c).
These instants are accompanied by the same chromophore index $n_{jl}$ [see Eq.~\eqref{Eq:H_S_B}] and the following superoperators [cf. Eq.~\eqref{Eq:def_Phi_t}]:
\begin{equation}
\label{Eq:super_s_l}
    s_l\leftrightarrow V_{n_{jl}}^\times,
\end{equation}
\begin{equation}
\label{Eq:super_s_j}
    s_j\leftrightarrow C_{n_{jl}}^r(s_l-s_j)V_{n_{jl}}^\times+iC_{n_{jl}}^i(s_l-s_j)V_{n_{jl}}^\circ.
\end{equation}
Having placed all the superoperators in a chronologically ordered string (reading diagrams from left to right), one performs integrations $\int_0^t ds_{2k}\dots\int_0^{s_2}ds_1$ over $2k$ internal instants and summations over $k$ independent chromophore indices.
Any diagram is either reducible or irreducible, depending on whether it can be cut into two by cutting a free Green's superoperator line or not.~\cite{Rammer_2007}
Examples of irreducible diagrams in Fig.~\ref{Fig:diags_final}(d) are diagrams (2), (4.2), and (4.3), while the diagram (4.1) is reducible.
Amputating the external lines corresponding to $\mathcal{G}_S(s_1)$ and $\mathcal{G}_S(t-s_{2k})$ of irreducible diagrams, one obtains the diagrammatic representation of the so-called (retarded) self-energy superoperator; see Fig.~\ref{Fig:diags_final}(f).
The self-energy superoperator thus consists of all amputated diagrams that cannot be cut into two by cutting a free Green's superoperator line.
One can then derive the following Dyson equation:~\cite{Rammer_2007}
\begin{equation}
\label{Eq:Dyson_eq_1}
\begin{split}
    &\mathcal{G}(t)=\mathcal{G}_S(t)\\&+\int_{-\infty}^{+\infty} ds_2\int_{-\infty}^{+\infty}ds_1\:\mathcal{G}_S(t-s_2)\Sigma(s_2-s_1)\mathcal{G}(s_1),
\end{split}
\end{equation}
whose diagrammatic representation is shown in Fig.~\ref{Fig:diags_final}(e).
All the superoperators in Eq.~\eqref{Eq:Dyson_eq_1} are retarded, and the integrals in Eq.~\eqref{Eq:Dyson_eq_1} can run from $-\infty$ to $+\infty$, which facilitates the transition to the frequency domain.
Rewriting Eq.~\eqref{Eq:Dyson_eq_1} as an equation for $\mathcal{U}(t)$, inserting the result thus obtained into Eq.~\eqref{Eq:def_dynamical_map}, and differentiating with respect to $t$, one obtains the QME,
\begin{equation}
\label{Eq:NZ-QME}
    \partial_t|\rho(t)\rrangle=-i\mathcal{L}_S|\rho(t)\rrangle-\int_0^t ds\:\mathcal{K}(t-s)|\rho(s)\rrangle,
\end{equation}
where the relation between the memory kernel $\mathcal{K}$ and the retarded self-energy $\Sigma$ is
\begin{equation}
\label{Eq:self-energy-memory-kernel}
    \Sigma(t)=-i\theta(t)\mathcal{K}(t).
\end{equation}

Equation~\eqref{Eq:NZ-QME} differs from the standard QME only by the representation of the memory kernel.
The memory kernel is commonly expressed in terms of projection superoperators $\mathcal{P}$ and $\mathcal{Q}=\mathbbm{1}-\mathcal{P}$, and its perturbation expansion in the exciton--environment interaction is most often hidden in the propagator of the irrelevant part $\mathcal{Q}\mathcal{L}_{S-B}\mathcal{Q}$ of the interaction Liouvillian, see, for example, Refs.~\onlinecite{Breuer-Petruccione-book,ChinJChemPhys.34.497}.
Choosing $\mathcal{P}|O\rrangle\leftrightarrow\mathrm{Tr}_B\{O\rho_B^\mathrm{eq}\}$~\cite{JChemPhys.55.2613,JChemPhys.62.4687} and expanding the aforementioned irrelevant propagator in power series, one obtains the perturbation expansion of the memory kernel in terms of the so-called partial cumulants of the interaction Liouvillian.~\cite{JPhysSocJpn.49.891}
The results of Ref.~\onlinecite{JChemPhys.62.4687} suggest that the same final result is obtained using the more conventional Agyres--Kelley~\cite{PhysRev.134.A98} projection superoperator $\mathcal{P}|O\rrangle\leftrightarrow\rho_B^\mathrm{eq}\mathrm{Tr}_B\{O\}$. 
One can convince themselves that the partial cumulants in Ref.~\onlinecite{JPhysSocJpn.49.891} simply restate that the memory-kernel (or self-energy) perturbation series does not contain disconnected diagrams, as in Fig.~\ref{Fig:diags_final}(f).
To that end, the general expressions of Ref.~\onlinecite{JPhysSocJpn.49.891} have to be transformed by applying Wick's theorem and diagrammatically representing the resulting series according to the above-stated rules.
These two steps, which were not considered in Ref.~\onlinecite{JPhysSocJpn.49.891}, are vital to the formulation of the self-consistent resummation scheme in Sec.~\ref{SSec:SCBA}. 

\subsection{Frequency-domain description}
Resummation techniques usually require transferring to the frequency space.
This is most conveniently done starting from Eq.~\eqref{Eq:Dyson_eq_1} and forming the dynamical equation for $\mathcal{G}(t)$,
\begin{equation}
\label{Eq:eom-retarded-G}
    \partial_t\mathcal{G}(t)=-i\delta(t)-i\mathcal{L}_S\mathcal{G}(t)-i\int_{-\infty}^{+\infty}ds\:\Sigma(t-s)\mathcal{G}(s).
\end{equation}
With the standard definition of the frequency-dependent quantities,
\begin{equation}
\label{Eq:G-t-Fourier-analyzed}
    \mathcal{G}(t)=\int_{-\infty}^{+\infty}\frac{d\omega}{2\pi}\:e^{-i\omega t}\mathcal{G}(\omega),
\end{equation}
Eq.~\eqref{Eq:eom-retarded-G} becomes the following algebraic equation:
\begin{equation}
\label{Eq:Dyson-freq}
    \mathcal{G}(\omega)=\left[\omega-\mathcal{L}_S-\Sigma(\omega)\right]^{-1}.
\end{equation}

The Hermitian property of the RDM implies that $\mathcal{G}(\omega)$ satisfies
\begin{equation}
\label{Eq:symmetry-property}
 \llangle e'_2 e'_1|\mathcal{G}(\omega)|e_2e_1\rrangle=-\llangle e'_1e'_2|\mathcal{G}(-\omega)|e_1e_2\rrangle^*.
\end{equation}
This property shows that it is sufficient to compute $\mathcal{G}(\omega)$ only for $\omega\geq 0$, while the values for negative frequencies follow from this symmetry property.
We additionally note that the same symmetry property also characterizes the inverse Green's superoperator $\mathcal{G}^{-1}(\omega)$, as well as the self-energy superoperator $\Sigma(\omega)$, which now carries the dimension of energy.

The frequency-domain diagrammatic representations of $\mathcal{G}(\omega)$ and $\Sigma(\omega)$ appear the same as the time-domain representations in Figs.~\ref{Fig:diags_final}(d) and~\ref{Fig:diags_final}(f), respectively.
The general rules for translating diagrams into formulas can be inferred from the discussion in Sec.~\ref{SSec:Born-Redfield}.

\subsection{Born and Redfield approximations}
\label{SSec:Born-Redfield}
\begin{figure*}[htbp!]
    \centering
    \includegraphics[width=0.9\textwidth]{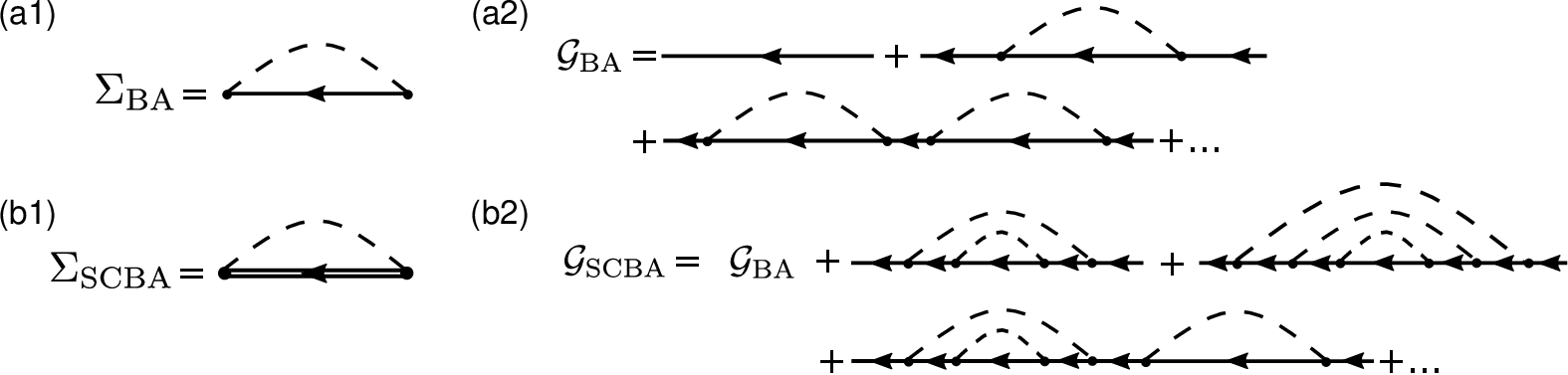}
    \caption{Diagrammatic representation of (a1) the BA self-energy superoperator [Eq.~\eqref{Eq:BA-self-energy-time} or~\eqref{Eq:BA-self-energy-freq}], (b1) the SCBA self-energy superoperator [Eq.~\eqref{Eq:SCBA-self-energy-freq}], (a2) Green's superoperator in the BA, and (b2) Green's superoperator in the SCBA.}
    \label{Fig:ba_scba_final}
\end{figure*}

The lowest, second-order approximation to $\Sigma$ is known as the (second) BA.~\cite{May-Kuhn-book}
The corresponding self-energy superoperator, shown in Fig.~\ref{Fig:ba_scba_final}(a1), reads as (see also Refs.~\onlinecite{JChemPhys.132.194111,PhysRevA.96.032105})
\begin{equation}
    \label{Eq:BA-self-energy-time}
 \Sigma_\mathrm{BA}(t)=\sum_nV_n^\times\mathcal{G}_S(t)\left[C_n^r(t)V_n^\times+iC_n^i(t)V_n^\circ\right],
\end{equation}
while its frequency-domain counterpart is
\begin{equation}
\label{Eq:BA-self-energy-freq}
\begin{split}
 &\Sigma_\mathrm{BA}(\omega)=\sum_n\int_{-\infty}^{+\infty}\frac{d\nu}{2\pi}\mathcal{J}_n(\omega-\nu)\\&\left\{\coth\left(\frac{\beta(\omega-\nu)}{2}\right)V_n^\times\mathcal{G}_S(\nu)V_n^\times+V_n^\times\mathcal{G}_S(\nu)V_n^\circ\right\}.
\end{split}
\end{equation}

Upon instering $\mathcal{K}_\mathrm{BA}(t)=i\Sigma_\mathrm{BA}(t)$ into Eq.~\eqref{Eq:NZ-QME}, we obtain the well-known TC2 QME,~\cite{PhysRevE.86.011915}
\begin{equation}
\label{Eq:tc2-qme}
\begin{split}
    &\partial_t\rho(t)=-i[H_S,\rho(t)]\\&-\sum_n\left[V_n,\int_0^t ds\:C_n(t-s)e^{-iH_S(t-s)}V_n\rho(s)e^{iH_S(t-s)}\right]\\&+\sum_n\left[V_n,\int_0^t ds\:C_n(t-s)^*e^{-iH_S(t-s)}\rho(s)V_ne^{iH_S(t-s)}\right].
\end{split}
\end{equation}
For delta-correlated noise characterized by the dephasing rate $\Gamma$, $C_n(t)=\Gamma\delta(t)$, Eq.~\eqref{Eq:tc2-qme} assumes the Lindblad form,
\begin{equation}
\label{Eq:tc2-qme-delta-correlated}
\begin{split}
    \partial_t\rho(t)=&-i[H_S,\rho(t)]\\&-\Gamma\sum_n\left(\frac{1}{2}\{V_n^2,\rho(t)\}-V_n\rho(t)V_n\right).
\end{split}
\end{equation}
We will use this result in Sec.~\ref{SSSec:only-phonons-vary-lambda-gamma-T}.

The time-independent and nonsecular Redfield theory (see, for example, Sec.~3.8.2 of Ref.~\onlinecite{May-Kuhn-book} or Sec.~11.2 of Ref.~\onlinecite{Valkunas-Abramavicius-Mancal-book})
\begin{equation}
\label{Eq:Redfield_equation}
    \partial_t|\rho(t)\rrangle=-i\mathcal{L}_S|\rho(t)\rrangle-\mathcal{R}|\rho(t)\rrangle,
\end{equation}
is obtained by inserting Eq.~\eqref{Eq:BA-self-energy-time} into Eq.~\eqref{Eq:eom-retarded-G} and then performing the Markovian and adiabatic approximations.
These approximations result in the delta-like self-energy in the time domain, $\Sigma_\mathrm{Red}(t)=\Sigma_\mathrm{Red}\delta(t)$, the frequency-independent self-energy $\Sigma_\mathrm{Red}(\omega)=\Sigma_\mathrm{Red}$, and the Redfield tensor $\mathcal{R}=i\Sigma_\mathrm{Red}$ [see Eq.~\eqref{Eq:self-energy-memory-kernel}], where 
\begin{equation}
\label{Eq:Sigma-Red}
    \Sigma_\mathrm{Red}=\int_{-\infty}^{+\infty}ds\:\Sigma_\mathrm{BA}(s)\:e^{i\mathcal{L}_Ss}.
\end{equation}
The Redfield theory is formulated in the exciton basis $\{|x\rangle\}$ defined through $H_S|x\rangle=\omega_x|x\rangle$.
Equation~\eqref{Eq:Sigma-Red} then implies that the Redfield-tensor matrix elements read 
\begin{equation}
\label{Eq:Sigma_Red_from_BA}
    \llangle x'_2x'_1|\mathcal{R}|x_2x_1\rrangle=i\llangle x'_2x'_1|\Sigma_\mathrm{BA}(\omega_{x_2}-\omega_{x_1})|x_2x_1\rrangle.
\end{equation}
For delta-correlated noise, $C_n(t)=\Gamma\delta(t)$, the Redfield equation [Eq.~\eqref{Eq:Redfield_equation}] also assumes the Lindblad form [Eq.~\eqref{Eq:tc2-qme-delta-correlated}].~\cite{Valkunas-Abramavicius-Mancal-book}

\subsection{Self-consistent Born approximation}
\label{SSec:SCBA}
In this work, we improve upon the BA by replacing the free-exciton Green's superoperator $\mathcal{G}_S$ in Fig.~\ref{Fig:ba_scba_final}(a1) or Eq.~\eqref{Eq:BA-self-energy-freq} with the interacting Green's superoperator $\mathcal{G}$; see Fig.~\ref{Fig:ba_scba_final}(b1).
The resultant equation for the self-energy superoperator
\begin{equation}
\label{Eq:SCBA-self-energy-freq}
\begin{split}
 &\Sigma(\omega)=\sum_n\int_{-\infty}^{+\infty}\frac{d\nu}{2\pi}\mathcal{J}_n(\omega-\nu)\\&\left\{\coth\left(\frac{\beta(\omega-\nu)}{2}\right)V_n^\times\mathcal{G}(\nu)V_n^\times+V_n^\times\mathcal{G}(\nu)V_n^\circ\right\}
\end{split}
\end{equation}
is to be solved together with the Dyson equation [Eq.~\eqref{Eq:Dyson-freq}] in a self-consistent loop.
Namely, one starts from the free-exciton case, $\Sigma^{(0)}(\omega)=0$, when Eq.~\eqref{Eq:Dyson-freq} gives the free-exciton Green's function ($\eta\to+0$),
\begin{equation}
\label{Eq:def-G-0th}
    \mathcal{G}^{(0)}(\omega)=\mathcal{G}_S(\omega)=[\omega+i\eta-\mathcal{L}_S]^{-1}.
\end{equation}
$\mathcal{G}^{(0)}(\omega)$ is then inserted into Eq.~\eqref{Eq:SCBA-self-energy-freq} to yield $\Sigma^{(1)}(\omega)=\Sigma_\mathrm{BA}(\omega)$, which is then inserted into Eq.~\eqref{Eq:Dyson-freq} to yield $\mathcal{G}^{(1)}(\omega)$, etc.
The above-described procedure is repeated until the difference between two consecutive iterations $\Sigma^{(k)}(\omega)$ and $\Sigma^{(k-1)}(\omega)$ for the self-energy ($k\geq 1$) becomes smaller than a prescribed numerical tolerance (see Sec.~\ref{SSec:technical} for more details).

Using either BA or SCBA for the self-energy superoperator, we do perform a partial resummation of the perturbation series for $\mathcal{G}$ [Fig.~\ref{Fig:diags_final}(d)].
The resulting $\mathcal{G}_\mathrm{(SC)BA}$ contains contributions from all orders in the exciton--environment interaction constant.
Nevertheless, the diagrammatic content of the SCBA is much richer than that of the BA, compare Fig.~\ref{Fig:ba_scba_final}(b2) to~\ref{Fig:ba_scba_final}(a2), suggesting that the SCBA is reliable in a much wider parameter range than the BA.
Still, the SCBA retains only a very limited subset of all possible diagrams appearing in Fig.~\ref{Fig:diags_final}(d),~\cite{PhysRevB.74.245104} and its reliability is to be carefully checked.

\section{Results}
\label{Sec:results}
\subsection{Technical details}
\label{SSec:technical}
Equation~\eqref{Eq:def-G-0th} features an infinitesimally small frequency $\eta$, which ensures the causality, i.e., $\mathcal{G}_S(t)=0$ for $t<0$.
We always shift $\omega\to\omega+i\eta$ on the right-hand side of the Dyson equation [Eq.~\eqref{Eq:Dyson-freq}], which, apart from the causality, ensures that the matrix inversion in Eq.~\eqref{Eq:Dyson-freq} is numerically stable.
However, the numerical Fourier transformation of $\mathcal{G(\omega)}$ thus obtained produces the exponentially damped Green's superoperator $\widetilde{\mathcal{G}}(t)=e^{-\eta t}\mathcal{G}(t)$.
The results for the true Green's superoperator $\mathcal{G}(t)=e^{\eta t}\widetilde{\mathcal{G}}(t)$ are thus the most reliable for $t\ll\eta^{-1}$.
In all our computations, we set $\eta=1\:\mathrm{cm}^{-1}$, meaning that our results for exciton dynamics are bound to be reliable for $t\ll 5\:\mathrm{ps}$.
While we find that our results in the real-time domain are free of finite-$\eta$ effects on timescales beyond $\eta^{-1}$, we always show only the initial 2--3\:ps of exciton dynamics.

Generally, $\mathcal{G}(\omega)$ slowly decays towards zero as $|\omega|\to+\infty$.
The dominant component of the high-frequency tail of $\mathcal{G}(\omega)$ can be inferred by performing one partial integration of $\mathcal{G}(\omega)=\int_0^{+\infty}dt\:e^{i(\omega+i\eta)t}\mathcal{G}(t)$, which results in
\begin{equation}
\label{Eq:G-omega-HF-tail}
    \mathcal{G}(\omega)=\frac{\mathbbm{1}}{\omega+i\eta}+\mathcal{O}(\omega^{-2}),\quad|\omega|\to+\infty.
\end{equation}
Deriving Eq.~\eqref{Eq:G-omega-HF-tail}, we use $\mathcal{G}(t=0)=-i\mathbbm{1}$, where $\mathbbm{1}$ denotes the unit operator in the Liouville space.
The strongly pronounced high-frequency tail of $\mathcal{G}(\omega)$ means that we have to consider many values of $\omega$ in order for the discrete Fourier transform to produce decent results in the time domain.
This is to be avoided as computing $\mathcal{G}(\omega)$ and $\Sigma(\omega)$ involves inversion of an $N^2\times N^2$ matrix [Eq.~\eqref{Eq:Dyson-freq}] and numerical integration [Eq.~\eqref{Eq:SCBA-self-energy-freq}], respectively.
Defining $\mathcal{T}(\omega)=\frac{\mathbbm{1}}{\omega+i\eta}$ and $\mathcal{G}^\mathrm{nt}(\omega)=\mathcal{G}(\omega)-\mathcal{T}(\omega)$,~\cite{mitric-thesis} the discrete (numerical) Fourier transformation of the non-tailed part $\mathcal{G}^\mathrm{nt}(\omega)$ produces $\widetilde{\mathcal{G}}^\mathrm{nt}(t)$, while the Fourier transformation of the high-frequency tail $\mathcal{T}(\omega)$ can be performed analytically to yield $\widetilde{\mathcal{T}}(t)=-i\theta(t)e^{-\eta t}\mathbbm{1}$.
Finally,
\begin{equation}
    \mathcal{G}(t)=-i\theta(t)\mathbbm{1}+e^{\eta t}\widetilde{\mathcal{G}}^\mathrm{nt}(t).
\end{equation}

As we assume local exciton--environment interaction [Eq.~\eqref{Eq:H_S_B}], $\mathcal{G}(\omega)$ and $\Sigma(\omega)$ are most conveniently represented in the site basis $\{|n\rangle\}$.
In all the examples to be discussed, we assume for simplicity that the environments of individual chromophores are identical.
The integration in Eq.~\eqref{Eq:SCBA-self-energy-freq} expressing  $\Sigma$ in terms of $\mathcal{G}$ is performed using the ordinary trapezoidal rule.
Because of the symmetry property in Eq.~\eqref{Eq:symmetry-property}, we perform a numerical integration of Eq.~\eqref{Eq:SCBA-self-energy-freq} from 0 to $\omega_\mathrm{max}$ with frequency step $\Delta\omega$.
The contribution around $\nu=\omega$ to the integral in Eq.~\eqref{Eq:SCBA-self-energy-freq} may be divergent for sub-Ohmic SDs, while it is finite (zero) for Ohmic (super-Ohmic) SDs.
The correct treatment of the possible divergence when computing Eq.~\eqref{Eq:SCBA-self-energy-freq} for sub-Ohmic SDs is beyond the scope of this study.
Importantly, the SDs that are most widely used in studying EET through multichromophoric aggregates are either super-Ohmic or Ohmic~\cite{JPhysChemB.117.7317,JChemPhys.141.094101} and can be handled in the manner presented here; see Sec.~\ref{SSec:fmo} and Appendix~\ref{App:FMO-details}.
We stop the self-consistent cycle once we achieve $\delta\Sigma^{(k)}\leq\varepsilon_\mathrm{tol}$, where ($k\geq 1$),
\begin{equation}
\label{Eq:def-delta-sigma-k}
 \delta\Sigma^{(k)}=\max_{\substack{n'_2n'_1\\n_2n_1}}\left|\frac{2\llangle n'_2n'_1|\Sigma^{(k)}(\omega)-\Sigma^{(k-1)}(\omega)|n_2n_1\rrangle}{\llangle n'_2n'_1|\Sigma^{(k)}(\omega)+\Sigma^{(k-1)}(\omega)|n_2n_1\rrangle}\right|,
\end{equation}
while $\varepsilon_\mathrm{tol}$ is the desired (relative) accuracy.
We summarize the values of adjustable numerical parameters ($\eta,\omega_\mathrm{max},\Delta\omega,\varepsilon_\mathrm{tol}$) involved in our computations in Table~\ref{Tab:numerical_parameters}.
% add new paragraph

The computational performance of the numerical integration of Eq.~\eqref{Eq:SCBA-self-energy-freq} mainly depends on the high-frequency rather than low-frequency behavior of the SD.
In the examples analyzed in Sec.~\ref{SSec:numerics_asymmetric_dimer}, $\mathcal{J}(\omega)\sim\omega^{-1}$ as $|\omega|\to+\infty$ so that the integrand in Eq.~\eqref{Eq:SCBA-self-energy-freq} falls off as $\nu^{-2}$ as $|\nu|\to+\infty$; see also Eq.~\eqref{Eq:G-omega-HF-tail}.
Other widely used SDs decrease even more rapidly in the high-frequency limit.~\cite{JPhysChemB.117.7317,JChemPhys.141.094101}
This suggests that the bottleneck in the self-consistent cycle is the inversion of the $N^2\times N^2$ matrix in Eq.~\eqref{Eq:Dyson-freq}.

\begin{figure}[htbp!]
    \centering
    \includegraphics[scale=1.0]{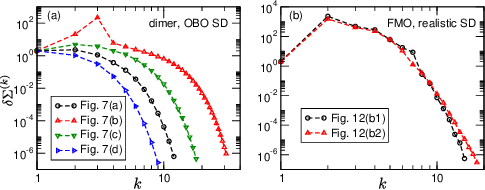}
    \caption{Dependence of the quantity $\delta\Sigma^{(k)}$ [Eq.~\eqref{Eq:def-delta-sigma-k}] monitoring the convergence of the self-consistent algorithm on the iteration number $k$ in the examples analyzed in (a) Figs.~\ref{Fig:Fig_scba_ph}(a)--\ref{Fig:Fig_scba_ph}(c), and (b) Figs.~\ref{Fig:fmo_pops}(b1) and~\ref{Fig:fmo_pops}(b2).
    Note the logarithmic scale on both axes.}
    \label{Fig:step_vs_error_090824}
\end{figure}

Figures~\ref{Fig:step_vs_error_090824}(a) and~\ref{Fig:step_vs_error_090824}(b), respectively, show $\delta\Sigma^{(k)}$ as a function $k$ when the SCBA is used on the dimer in the overdamped phonon continuum [the examples analyzed in Figs.~\ref{Fig:Fig_scba_ph}(a)--\ref{Fig:Fig_scba_ph}(d)] and on the seven-site FMO model in the realistic environment [the examples analyzed in Figs.~\ref{Fig:fmo_pops}(b1) and~\ref{Fig:fmo_pops}(b2)].
Notably, the convergence of the self-consistent algorithm is achieved in a couple of tens of steps, even in a multichromophoric system immersed in a structured environment.
After its initial increase with $k$, starting from the value of 2 [see the text above Eq.~\eqref{Eq:def-G-0th}], $\delta\Sigma^{(k)}$ decreases in a power-law fashion for sufficiently large $k$.

\begin{table}[htbp!]
    \centering
    \begin{tabular}{c|c}
        Parameter (unit) & Value\\\hline
        $\eta$ ($\mathrm{cm}^{-1}$) & 1\\
        $\omega_\mathrm{max}$ ($\mathrm{cm}^{-1}$) & $3000-5000$\\
        $\Delta\omega$ ($\mathrm{cm}^{-1}$) & 0.5\\
        $\varepsilon_\mathrm{tol}$ (-) & $10^{-6}$
    \end{tabular}
    \caption{Summary of the adjustable numerical parameters needed to perform the self-consistent cycle and the values used in our computations.}
    \label{Tab:numerical_parameters}
\end{table}

In both BA and SCBA, the trace of the RDM is preserved because of the outermost commutator in self-energies in Eqs.~\eqref{Eq:BA-self-energy-freq} and~\eqref{Eq:SCBA-self-energy-freq}.
The RDM positivity is a much subtler issue, but we observe that, whenever SCBA improves over the BA, the results of both approximations conform to the positivity requirement on the timescales analyzed.

As we are primarily interested in examining the reliability of the BA and SCBA, we use
\begin{equation}
\label{Eq:init-cond-eff-avg}
    |\rho(0)\rrangle=\frac{1}{N}\sum_{n_2n_1}|n_2n_1\rrangle
\end{equation}
as the initial condition in all numerical computations.
Apart from its simplicity, this initial condition provides a fairer assessment of the approximation performance than the widely used initial condition $|n_0n_0\rrangle$, in which the exciton is placed at chromophore $n_0$.
In more detail, exciton dynamics in an arbitrary basis $\{|e\rangle\}$ is
\begin{equation}
\label{Eq:EET-dynamics-eff-avg}
\begin{split}
    &\llangle e_2e_1|\rho(t)\rrangle=\sum_{\substack{n_2n_1\\n'_2n'_1}}\langle e_2|n_2\rangle\langle n_1|e_1\rangle\times\\&\llangle n_2n_1|\mathcal{G}(t)|n'_2n'_1\rrangle\llangle n'_2n'_1|\rho(0)\rrangle.
\end{split}
\end{equation}
If the exciton is initially placed at chromophore $n_0$, its subsequent dynamics is determined by only $N^2$ matrix elements $\llangle n_2n_1|\mathcal{G}(t)|n_0n_0\rrangle$ of $\mathcal{G}$ out of the total of $N^4$ elements.
Quite generally,~\cite{SciRep.6.28204} the quality of approximate dynamics is different for different matrix elements of $\mathcal{G}$, that is, for different starting chromophores $n_0$. 
Inserting the initial condition of Eq.~\eqref{Eq:init-cond-eff-avg} into Eq.~\eqref{Eq:EET-dynamics-eff-avg}, we can assess the overall approximation performance, which is effectively "averaged" over different matrix elements of $\mathcal{G}$.

\subsection{Reliability of the SCBA: Asymmetric dimer}
\label{SSec:numerics_asymmetric_dimer}
The advantages and shortcomings of the SCBA are most transparently identified on the simplest model relevant for EET, the molecular dimer.
In the site basis $\{|1_n\rangle,|2_n\rangle\}$, the exciton Hamiltonian $H_S$ is represented by the matrix 
\begin{equation}
\label{Eq:def-H_S-dimer}
    H_S=\begin{pmatrix}
        \Delta\varepsilon & J\\
        J & 0
    \end{pmatrix},
\end{equation}
where $\Delta\varepsilon$ is the site-energy gap, while $J$ is the resonance coupling. 
The exciton state of lower (higher) energy is denoted as $|1_x\rangle$ ($|2_x\rangle$).

In Secs.~\ref{SSSec:only-phonons}--\ref{SSSec:only-phonons-self-energy}, we consider exciton dynamics in the featureless phonon environment described by the overdamped Brownian oscillator (OBO) SD,~\cite{Mukamel-book} 
\begin{equation}
\label{Eq:def-OBO-SD}
\mathcal{J}_\mathrm{ph}(\omega)=2\lambda_\mathrm{ph}\frac{\omega\gamma_\mathrm{ph}}{\omega^2+\gamma_\mathrm{ph}^2},
\end{equation}
where $\lambda_\mathrm{ph}$ is the reorganization energy, while $\gamma_\mathrm{ph}^{-1}$ determines the environment-reorganization time scale.
Table~\ref{Tab:model_parameters} summarizes the default values of model parameters, which are broadly representative of photosynthetic aggregates.
The performance of our approximations upon varying these parameters is studied in Secs.~\ref{SSSec:only-phonons} and~\ref{SSSec:only-phonons-vary-lambda-gamma-T}. 
\begin{table}
    \centering
    \begin{tabular}{c|c}
        Parameter (unit) & Value\\\hline
        $\Delta\varepsilon$ ($\mathrm{cm}^{-1}$) & 100\\
        $J$ ($\mathrm{cm}^{-1}$) & 50\\
        $\lambda_\mathrm{ph}$ ($\mathrm{cm}^{-1}$) & 40\\
        $\gamma_\mathrm{ph}$ ($\mathrm{cm}^{-1}$) & 40\\
        $T$ (K) & 300
    \end{tabular}
    \caption{Summary of the default values of excitonic ($\Delta\varepsilon,J$) and exciton--environment interaction ($\lambda_\mathrm{ph},\gamma_\mathrm{ph}$) parameters used in Secs.~\ref{SSSec:only-phonons}--\ref{SSSec:only-phonons-self-energy}.}
    \label{Tab:model_parameters}
\end{table}

In Sec.~\ref{SSSec:one-underdamped-vib}, we assume that the exciton interacts with a single underdamped vibrational mode so that the interaction can be modeled by the underdamped Brownian oscillator SD~\cite{Mukamel-book}
\begin{equation}
\label{Eq:def-UBO-SD}
\mathcal{J}_\mathrm{vib}(\omega)=S_0\omega_0\left[\frac{\omega\gamma_0}{(\omega-\omega_0)^2+\gamma_0^2}+\frac{\omega\gamma_0}{(\omega+\omega_0)^2+\gamma_0^2}\right],
\end{equation}
where $\omega_0$ is the vibrational frequency, $S_0$ the Huang--Rhys factor, while $\gamma_0$ is the relaxation rate.
Quite generally, $\gamma_0\ll\omega_0$ and $S_0\ll 1$.
The values used in benchmarks are taken from Ref.~\onlinecite{JChemPhys.157.095103} and are listed in Sec.~\ref{SSSec:one-underdamped-vib}.

\subsubsection{Overdamped phonon continuum: Variations in excitonic parameters}
\label{SSSec:only-phonons}

\begin{figure*}[htbp!]
    \centering
    \includegraphics[width=\textwidth]{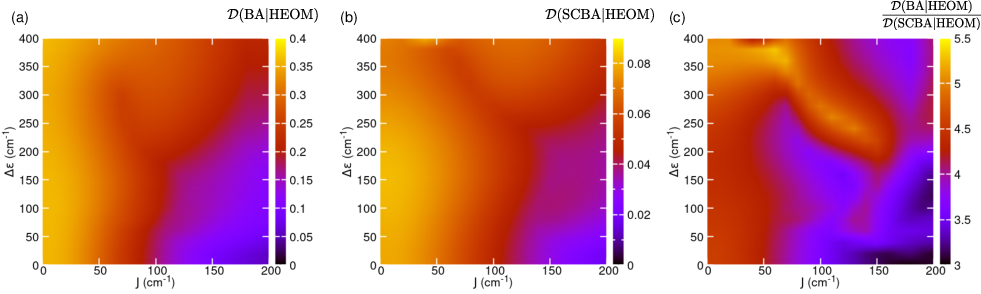}
    \caption{(a) and (b): Heat maps of the maximum trace distance $\mathcal{D}(\mathrm{(SC)BA|HEOM})$ [Eq.~\eqref{Eq:def-trace-distance}] between the numerically exact and BA [(a)] or SCBA [(b)] excitonic RDM over the interval $[0,t_\mathrm{max}]$.
    (c): Heat map of the ratio $\frac{\mathcal{D}(\mathrm{BA|HEOM})}{\mathcal{D}(\mathrm{SCBA|HEOM})}$ between the performance metrics used in (a) and (b).
    All quantities are computed for different values of the resonance coupling $J$ and the site-energy gap $\Delta\varepsilon$, the remaining parameters assume their default values listed in Table~\ref{Tab:model_parameters}, the initial condition is specified in Eq.~\eqref{Eq:init-cond-eff-avg}, while $t_\mathrm{max}=2\:\mathrm{ps}$.}
    \label{Fig:dimer-obo-heat-J-de}
\end{figure*}

Here, we fix the exciton--environment interaction parameters $\lambda_\mathrm{ph}$ and $\gamma_\mathrm{ph}$, and study the quality of SCBA and BA for different excitonic parameters $J$ and $\Delta\varepsilon$.
Figure~\ref{Fig:dimer-obo-heat-J-de} provides an overall assessment of the performance of BA [Fig.~\ref{Fig:dimer-obo-heat-J-de}(a)] and SCBA [Fig.~\ref{Fig:dimer-obo-heat-J-de}(b)], and of the improvement of the SCBA
over the BA introduced by the self-consistent cycle
[Fig.~\ref{Fig:dimer-obo-heat-J-de}(c)].
A convenient performance measure is the trace distance between the approximate [$\rho_\mathrm{(SC)BA}(t)$] and numerically exact [$\rho_\mathrm{HEOM}(t)$] RDM.
As the trace distance is time-dependent, while we also limit ourselves to the short-time dynamics (see Sec.~\ref{SSec:technical}), we quantify the performance of our approximations using the maximum trace distance over the time window $[0,t_\mathrm{max}]$:
\begin{equation}
\label{Eq:def-trace-distance}
\begin{split}
    &\mathcal{D}(\mathrm{(SC)BA|HEOM})=\\&\max_{0\leq t\leq t_\mathrm{max}}\frac{1}{2}\sum_{k=1}^2 \left|r_k^\mathrm{(SC)BA-HEOM}(t)\right|,
\end{split}
\end{equation}
where $\displaystyle{r_k^\mathrm{(SC)BA-HEOM}(t)}$ is the $k$-th eigenvalue of the operator $\rho_\mathrm{(SC)BA}(t)-\rho_\mathrm{HEOM}(t)$.
We set $t_\mathrm{max}=2\:\mathrm{ps}$ in Eq.~\eqref{Eq:def-trace-distance}.
The colorbar ranges in Figs.~\ref{Fig:dimer-obo-heat-J-de}(a) and~\ref{Fig:dimer-obo-heat-J-de}(b) suggest that the SCBA is generally a better approximation to the exact dynamics than the BA.
To quantify the improvement of the SCBA
over the BA, in Fig.~\ref{Fig:dimer-obo-heat-J-de}(c) we plot the ratio $\frac{\mathcal{D}(\mathrm{BA|HEOM})}{\mathcal{D}(\mathrm{SCBA|HEOM})}$, which we find to be greater than or equal to unity for all the pairs $(J,\Delta\varepsilon)$ examined.
The larger the ratio, the more pronounced the improvement of the SCBA
over the BA.

Figures~\ref{Fig:dimer-obo-heat-J-de}(a) and~\ref{Fig:dimer-obo-heat-J-de}(b) show that the reliability of both BA and SCBA generally improves with increasing $J$ and/or decreasing $\Delta\varepsilon$.
This suggests that the approximations are best suited for relatively delocalized excitons, when the mixing angle $\theta\in[0,\pi/4]$, defined as $\tan(2\theta)=2J/\Delta\varepsilon$, significantly deviates from zero.
Still, for $\Delta\varepsilon=0$, when excitons are perfectly delocalized, the quality of both BA and SCBA increases with increasing $J$, i.e., decreasing $\lambda_\mathrm{ph}/J$.
The impact of the exciton--environment interaction on the approximation reliability will be analyzed in detail in Sec.~\ref{SSSec:only-phonons-vary-lambda-gamma-T}. 
Figure~\ref{Fig:dimer-obo-heat-J-de}(c) reveals that the improvement of SCBA
over the BA is the most pronounced in the region of moderate resonance coupling $50\:\mathrm{cm}^{-1}\lesssim J\lesssim 150\:\mathrm{cm}^{-1}$ and large site-energy gap $\Delta\varepsilon\gtrsim 200\:\mathrm{cm}^{-1}$.
The improvement is also appreciable for small resonance coupling, irrespective of the value of $\Delta\varepsilon$, when both BA and SCBA perform relatively poorly; see Figs.~\ref{Fig:dimer-obo-heat-J-de}(a) and~\ref{Fig:dimer-obo-heat-J-de}(b).

\subsubsection{Overdamped phonon continuum: Analytical insights into the pure-dephasing model}
\label{SSSec:only-phonons-analytical}

For $J=0$, the model reduces to the pure-dephasing model, in which there is no population dynamics, but only dephasing of the initially present interexciton coherences.
The superoperators entering Eq.~\eqref{Eq:def_Phi_t} are then time independent, and combining Eqs.~\eqref{Eq:U_t_FV}--\eqref{Eq:def-G-t}, we readily obtain the following exact expression for the reduced evolution superoperator: 
\begin{equation}
\label{Eq:exact_G_t_pure_phase_noise}
\begin{split}
    \llangle n'_2n'_1|\mathcal{G}(t)|n_2n_1\rrangle=-i\theta(t)\delta_{n'_2n_2}\delta_{n'_1n_1}\left\{\delta_{n_2n_1}+ \right. \\ \left.
    (1-\delta_{n_2n_1})e^{-i\varepsilon_{n_2n_1}t}e^{-2g^r(t)}\right\},
\end{split}
\end{equation}
where $\displaystyle{g^r(t)=\int_0^t ds_2\int_0^{s_2}ds_1\:C^r(s_1)}$ is the real part of the line shape function, whereas $\varepsilon_{n_2n_1}=\varepsilon_{n_2}-\varepsilon_{n_1}$ (for the dimer, $\varepsilon_{n_2n_1}=\pm\Delta\varepsilon$).
The derivation of Eq.~\eqref{Eq:exact_G_t_pure_phase_noise}, in which only $g^r$ appears, crucially relies on our assumption that individual-chromophore environments are identical. 
While the exact coherence dynamics, which follows from Eqs.~\eqref{Eq:exact_G_t_pure_phase_noise},~\eqref{Eq:EET-dynamics-eff-avg}, and~\eqref{Eq:init-cond-eff-avg}, can be recovered from the time-convolutionless second-order QME,~\cite{ChemPhys.347.243} the corresponding BA and SCBA results remain only approximations to the exact solution, as both involve an explicit convolution in the time domain.
Still, relevant analytical insights concerning the (SC)BA can be obtained for the pure-dephasing model in the high-temperature limit $2\pi T\gg\gamma_\mathrm{ph}$.
In Appendix~\ref{App:analytics}, we derive that the matrix elements of the exact self-energy superoperator $\llangle n_2n_1|\Sigma(\omega)|n_2n_1\rrangle$ for $n_2\neq n_1$ have the following continued-fraction expansion (CFE):
\begin{widetext}
\begin{equation}
\label{Eq:exact_CFE}
\begin{split}
    \llangle n_2n_1|\Sigma(\omega)|n_2n_1\rrangle=
    \cfrac{4\lambda_\mathrm{ph}T}{\omega-\varepsilon_{n_2n_1}+i\gamma_\mathrm{ph}-\cfrac{2\cdot 4\lambda_\mathrm{ph}T}{\omega-\varepsilon_{n_2n_1}+2i\gamma_\mathrm{ph}-\cfrac{3\cdot 4\lambda_\mathrm{ph}T}{\omega-\varepsilon_{n_2n_1}+3i\gamma_\mathrm{ph}-\cdots}}}.
\end{split}
\end{equation}
\end{widetext}
Truncating the CFE in the first layer, we obtain the BA result ($n_2\neq n_1$),
\begin{equation}
\label{Eq:BA_sigma_pure_phase_noise}
    \llangle n_2n_1|\Sigma_\mathrm{BA}(\omega)|n_2n_1\rrangle=\frac{4\lambda_\mathrm{ph}T}{\omega-\varepsilon_{n_2n_1}+i\gamma_\mathrm{ph}}.
\end{equation}
The same result follows from Eq.~\eqref{Eq:BA-self-energy-freq} upon approximating $\mathrm{coth}\left(\frac{\beta(\omega-\nu)}{2}\right)\approx\frac{2T}{\omega-\nu}$, as appropriate in the high-temperature limit, and performing a contour integration by closing the contour in the upper half-plane.
In the pure-dephasing model, we can use Eqs.~\eqref{Eq:Sigma_Red_from_BA} and~\eqref{Eq:BA_sigma_pure_phase_noise} to obtain the matrix elements of the self-energy superoperator in the Redfield theory,
\begin{equation}
\label{Eq:Redfield_Sigma}
    \llangle n_2n_1|\Sigma_\mathrm{Red}|n_2n_1\rrangle=-i\frac{4\lambda_\mathrm{ph}T}{\gamma_\mathrm{ph}}.
\end{equation}
Appendix~\ref{App:analytics} also demonstrates that the CFE of the SCBA self-energy reads
\begin{widetext}
\begin{equation}
\label{Eq:SCBA_CFE}
    \llangle n_2n_1|\Sigma_\mathrm{SCBA}(\omega)|n_2n_1\rrangle=\cfrac{4\lambda_\mathrm{ph}T}{\omega-\varepsilon_{n_2n_1}+i\gamma_\mathrm{ph}-\cfrac{4\lambda_\mathrm{ph}T}{\omega-\varepsilon_{n_2n_1}+2i\gamma_\mathrm{ph}-\cfrac{4\lambda_\mathrm{ph}T}{\omega-\varepsilon_{n_2n_1}+3i\gamma_\mathrm{ph}-\cdots}}}.
\end{equation}
\end{widetext}

\begin{figure}
    \centering
    \includegraphics[width=\columnwidth]{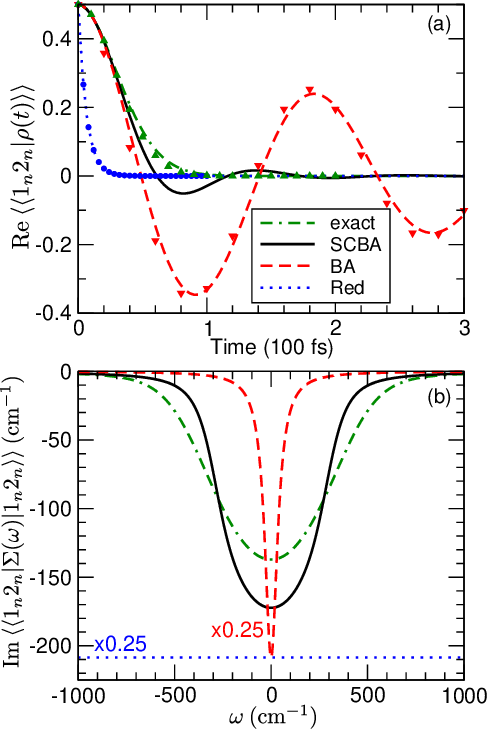}
    \caption{(a) Time dependence of the real part of the coherence in a symmetric ($\Delta\varepsilon=0$) pure-dephasing dimer.
    (b) Frequency profile of the imaginary part of the matrix element of the self-energy superoperator describing coherence dephasing in a symmetric pure-dephasing dimer.
    We compare the results of the SCBA (solid lines), BA (dashed lines), and Redfield theory (dotted lines) to the exact result (double dashed-dotted lines).
    The lines in (a) display Fourier-transformed frequency-domain results using the self-energies given in Eq.~\eqref{Eq:exact_CFE} (label "exact"), Eq.~\eqref{Eq:SCBA_CFE} (label "SCBA"), Eq.~\eqref{Eq:BA_sigma_pure_phase_noise} (label "BA"), and Eq.~\eqref{Eq:Redfield_Sigma} (label "Red").
    Full symbols in (a) display the analytical results given by Eq.~\eqref{Eq:coh-deph-Redfield} (circles), Eq.~\eqref{Eq:coh-deph-BA} (down-triangles), and Eq.~\eqref{Eq:coh-deph-exact} (up-triangles).
    For visual clarity, the BA and Redfield self-energies in (b) are scaled down by a factor of 4.
    }
    \label{Fig:figure_gamma_40cm-1}
\end{figure}

Comparing Eq.~\eqref{Eq:exact_CFE} to Eq.~\eqref{Eq:SCBA_CFE}, we find that the CFE of the SCBA result can be obtained from the exact result by changing all the coefficients multiplying $4\lambda_\mathrm{ph}T$ in the CFE numerators to unity.
While this could suggest that the SCBA is, in general, a poor approximation to the exact solution,~\cite{PhysRevB.56.4494,PhysRevB.74.245104} we note that the broadening factors are the same at each CFE denominator in both the SCBA and exact self-energy.
One can then expect that the SCBA is superior to both the BA and the Redfield theory at reproducing the timescale of coherence dephasing.
This expectation is confirmed in Fig.~\ref{Fig:figure_gamma_40cm-1}(a) comparing different approximations to the coherence dynamics in a symmetric ($\Delta\varepsilon=0$) pure-dephasing dimer.
The Redfield theory predicts an excessively fast coherence dephasing whose analytical form reads as [see Appendix~\ref{App:analytics} and the circles in Fig.~\ref{Fig:figure_gamma_40cm-1}(a)]
\begin{equation}
\label{Eq:coh-deph-Redfield}
\llangle 1_n2_n|\rho_\mathrm{Red}(t)\rrangle=\frac{1}{2}\exp\left(-\frac{4\lambda_\mathrm{ph}T}{\gamma_\mathrm{ph}}t\right).
\end{equation}
On the contrary, the BA predicts an excessively slow coherence dephasing that can be reasonably described by [see Appendix~\ref{App:analytics} and the down-triangles in Fig.~\ref{Fig:figure_gamma_40cm-1}(a)]
\begin{equation}
\label{Eq:coh-deph-BA}
\llangle 1_n2_n|\rho_\mathrm{BA}(t)\rrangle\approx\frac{1}{2}\cos\left(2\sqrt{\lambda_\mathrm{ph}T}\:t\right)e^{-\gamma_\mathrm{ph}t/2}.
\end{equation}
The exact coherence-dephasing timescale is in between the results of the Redfield theory and BA, while the corresponding exact dynamics can be reasonably approximated by [see Appendix~\ref{App:analytics} and the up-triangles in Fig.~\ref{Fig:figure_gamma_40cm-1}(a)]
\begin{equation}
\label{Eq:coh-deph-exact}
\llangle 1_n2_n|\rho(t)\rrangle\approx\frac{1}{2}\exp\left(-2\lambda_\mathrm{ph}Tt^2\right).
\end{equation}
The SCBA indeed reproduces the correct order of magnitude of the dephasing timescale, although it displays oscillations similar to that predicted by Eq.~\eqref{Eq:coh-deph-BA}.
The performance of different approximations can also be inferred from (the imaginary part of) the corresponding self-energy profile presented in Fig.~\ref{Fig:figure_gamma_40cm-1}(a).
The self-energy within the Redfield theory [Eq.~\eqref{Eq:Redfield_Sigma}] does not bear even a qualitative resemblance to the exact result.
The BA, SCBA, and the exact result all display a peak centered around $\omega=0$.
The BA peak is the narrowest and highest and has a Lorentzian shape whose full width at half-maximum is determined by $\gamma_\mathrm{ph}$ only; see Eq.~\eqref{Eq:BA_sigma_pure_phase_noise}.
The exact peak [Eq.~\eqref{Eq:exact_CFE}] is much broader than the BA peak, while the width of the SCBA peak [Eq.~\eqref{Eq:SCBA_CFE}] is somewhat smaller than, yet comparable to, the width of the exact peak.

\subsubsection{Overdamped phonon continuum: Variations in exciton--environment interaction parameters and temperature}
\label{SSSec:only-phonons-vary-lambda-gamma-T}

\begin{figure*}[htbp!]
    \centering
    \includegraphics[width=\textwidth]{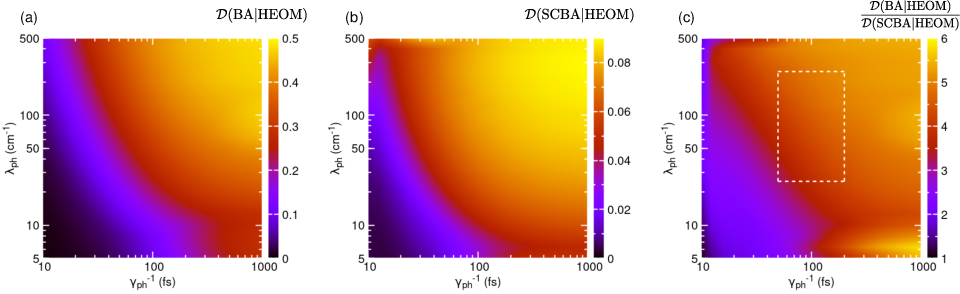}
    \caption{(a) and (b): Heat maps of the maximum trace distance $\mathcal{D}(\mathrm{(SC)BA|HEOM})$ between the numerically exact and BA [(a)] or SCBA [(b)] excitonic RDM over the interval $[0,t_\mathrm{max}]$ [Eq.~\eqref{Eq:def-trace-distance}].
    (c) Heat map of the ratio $\frac{\mathcal{D}(\mathrm{BA|HEOM})}{\mathcal{D}(\mathrm{SCBA|HEOM})}$ between the performance metrics used in (a) and (b).
    All quantities are computed for different values of the reorganization energy $\lambda_\mathrm{ph}$ and the environment-reorganization timescale $\gamma_\mathrm{ph}^{-1}$, the remaining parameters assume their default values listed in Table~\ref{Tab:model_parameters}, the initial condition is specified in Eq.~\eqref{Eq:init-cond-eff-avg}, while $t_\mathrm{max}=2\:\mathrm{ps}$.
    The dashed-line rectangle in (c) delimits the ranges of values of $\gamma_\mathrm{ph}^{-1}$ and $\lambda_\mathrm{ph}$ typically used in modeling photosynthetic EET.}
    \label{Fig:dimer-obo-heat}
\end{figure*}

Here, we fix exciton parameters $\Delta\varepsilon$ and $J$, and vary $\lambda_\mathrm{ph}$ and $\gamma_\mathrm{ph}$ to examine the reliability of BA [Fig.~\ref{Fig:dimer-obo-heat}(a)] and SCBA [Fig.~\ref{Fig:dimer-obo-heat}(b)], as well as the improvement over the BA brought about by the self-consistent cycle [Fig.~\ref{Fig:dimer-obo-heat}(c)].
Overall, we find that the quality of both BA and SCBA improves with decreasing the reorganization energy and/or shortening the environment-reorganization timescale; see Figs.~\ref{Fig:dimer-obo-heat}(a) and~\ref{Fig:dimer-obo-heat}(b).
Figure~\ref{Fig:dimer-obo-heat}(c) shows that the SCBA is always better at approximating the exact dynamics than the BA.
Within the range of values of $\lambda_\mathrm{ph}$ and $\gamma_\mathrm{ph}^{-1}$ typically used in models of photosynthetic EET ($25\:\mathrm{cm}^{-1}\lesssim\lambda_\mathrm{ph}\lesssim 250\:\mathrm{cm}^{-1}$ and $50\:\mathrm{fs}\lesssim\gamma_\mathrm{ph}^{-1}\lesssim 200\:\mathrm{fs}$),~\cite{RevModPhys.90.035003} the self-consistent cycle significantly improves the BA results; see the dashed-line rectangle in Fig.~\ref{Fig:dimer-obo-heat}(c).
Notably, for very fast environments ($\gamma_\mathrm{ph}^{-1}\lesssim 10\:\mathrm{fs}$), we find that the quality of BA and SCBA is virtually the same, i.e., $\mathcal{D}(\mathrm{SCBA|HEOM})\approx\mathcal{D}(\mathrm{BA|HEOM})$, for all reorganization energies examined.
To understand this, we observe that when $\gamma_\mathrm{ph}\to\infty$ and $T\to\infty$ so that $\gamma_\mathrm{ph}/T\to 0$, our model reduces to the well-known Haken--Strobl--Reineker white-noise model,~\cite{ZPhysA.262.135,ZPhys.249.253} within which $C_n(t)=\Gamma\delta(t)$, with $\Gamma=2\lambda_\mathrm{ph}T/\gamma_\mathrm{ph}$.
Inserting this $C_n(t)$ into Eq.~\eqref{Eq:def_Phi_t} for the exact evolution superoperator $\mathcal{U}(t)$, we conclude that the exact equation of motion for $\rho(t)$ coincides with the Lindblad-like equation~\eqref{Eq:tc2-qme-delta-correlated}. 
To reach this conclusion, we use that the white-noise assumption renders the time-ordering sign upon differentiating Eq.~\eqref{Eq:def_Phi_t} effective only on the superoperator $e^{-\Phi(t)}$ and ineffective on the superoperator $\partial_t\Phi(t)$.
Our discussion in Sec.~\ref{SSec:Born-Redfield} then implies that the BA, and even the Redfield theory, becomes exact in the white-noise limit.
The exactness of the BA implies that the self-consistent cycle cannot improve on BA any further, meaning that the SCBA result is also exact in the white-noise limit.

\begin{figure*}[htbp!]
    \centering
    \includegraphics[width=\textwidth]{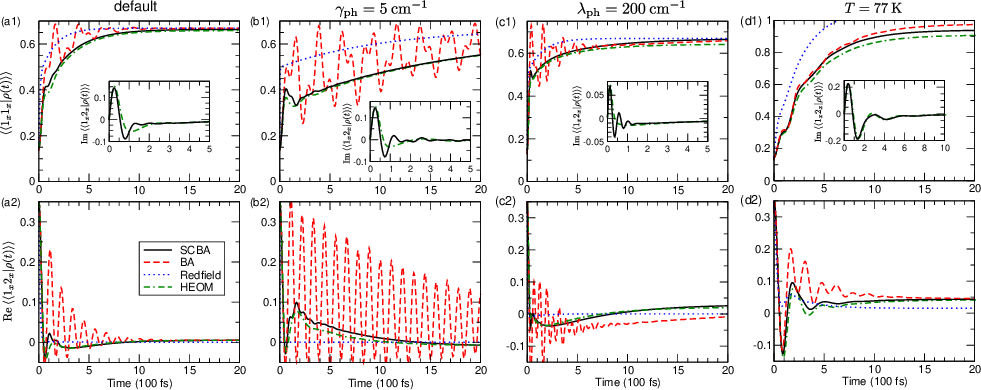}
    \caption{Dynamics of the population of the lower-energy exciton state $|1_x\rangle$ [(a1)--(d1)] and the real part of the interexciton coherence [(a2)--(d2)] computed using SCBA (solid lines), BA (dashed lines), Redfield theory (dotted lines), and HEOM (double dash-dotted lines).
    The insets in (a1)--(d1) present the very initial dynamics of the imaginary part of the interexciton coherence (BA and Redfield results are omitted for visual clarity).
    The parameters $\lambda_\mathrm{ph},\gamma_\mathrm{ph},T$ and $\lambda_\mathrm{ph}$ assume their default values in (a1) and (a2), while we change $\gamma_\mathrm{ph}=5\:\mathrm{cm}^{-1}$ in (b1) and (b2), $T=77\:\mathrm{K}$ in (c1) and (c2), and $\lambda_\mathrm{ph}=200\:\mathrm{cm}^{-1}$ in (d1) and (d2).}
    \label{Fig:Fig_scba_ph}
\end{figure*}

Figure~\ref{Fig:dimer-obo-heat}(c) shows that the improvement of the SCBA
over the BA is the most pronounced for sluggish environments or large reorganization energies. 
Both observations are ultimately rooted in the richer diagrammatic content of the SCBA compared to that of the BA, see Fig.~\ref{Fig:ba_scba_final}, and are remarkable as computing exciton dynamics modulated by strong exciton--environment interactions and/or slow environments has to take into account higher-order environmentally assisted processes.~\cite{JChemPhys.132.194111,PhysRevA.96.032105}
Similar challenges are encountered when studying exciton dynamics at low temperatures.
Additional insights into the quality of different approximations can be gained from Fig.~\ref{Fig:Fig_scba_ph}, which compares approximate and numerically exact exciton dynamics for the default values of model parameters [Figs.~\ref{Fig:Fig_scba_ph}(a1) and~\ref{Fig:Fig_scba_ph}(a2)], slow environment [$\gamma_\mathrm{ph}=5\:\mathrm{cm}^{-1}$, Figs.~\ref{Fig:Fig_scba_ph}(b1) and~\ref{Fig:Fig_scba_ph}(b2)], strong exciton--environment interaction [$\lambda_\mathrm{ph}=200\:\mathrm{cm}^{-1}$, Figs.~\ref{Fig:Fig_scba_ph}(c1) and~\ref{Fig:Fig_scba_ph}(c2)], and low temperature [$T=77\:\mathrm{K}$, Figs.~\ref{Fig:Fig_scba_ph}(d1) and~\ref{Fig:Fig_scba_ph}(d2)]. 
The overall performance of approximate methods is essentially as discussed in Sec.~\ref{SSSec:only-phonons-analytical}.
The Redfield theory shows pronounced deviations from the exact result already on shortest timescales, whereas the BA and SCBA reproduce the very initial stages ($t\lesssim 50\:\mathrm{fs}$) of the exact dynamics quite well.
While the subsequent dynamics within the BA generally exhibits oscillatory features that are damped relatively slowly, cf. Fig.~\ref{Fig:figure_gamma_40cm-1}(a), the SCBA dynamics of exciton populations and the interexciton coherence follows the corresponding HEOM results very reasonably.
The SCBA approximates the true exciton-population dynamics in both realistically slow [Fig.~\ref{Fig:Fig_scba_ph}(a1)] and excessively slow [Fig.~\ref{Fig:Fig_scba_ph}(b1)] environments quite well, both on subpicosecond and on somewhat longer timescales.
Meanwhile, Figs.~\ref{Fig:Fig_scba_ph}(a2) and~\ref{Fig:Fig_scba_ph}(b2), as well as the insets of Figs.~\ref{Fig:Fig_scba_ph}(a1) and~\ref{Fig:Fig_scba_ph}(b1), suggest that the SCBA is not that good at reproducing the true interexciton-coherence dynamics.
Still, it does capture the correct long-time behavior (the imaginary part of the interexciton coherence tends to zero, and the real part tends to a non-zero value).
For strong interactions and at low temperatures, the subpicosecond dynamics of exciton populations within the SCBA is quite close to the true dynamics, see Figs.~\ref{Fig:Fig_scba_ph}(c1) and~\ref{Fig:Fig_scba_ph}(d1), while some quantitative differences between them appear on longer time scales (these are more pronounced for the higher-energy exciton state).
Meanwhile, the SCBA dynamics of the interexciton coherence agrees well with the corresponding numerically exact dynamics, especially at lower temperatures; see Figs.~\ref{Fig:Fig_scba_ph}(c2) and~\ref{Fig:Fig_scba_ph}(d2) and the insets of Figs.~\ref{Fig:Fig_scba_ph}(c1) and~\ref{Fig:Fig_scba_ph}(d1).

\begin{figure}[htbp!]
    \centering
    \includegraphics[width=0.8\columnwidth]{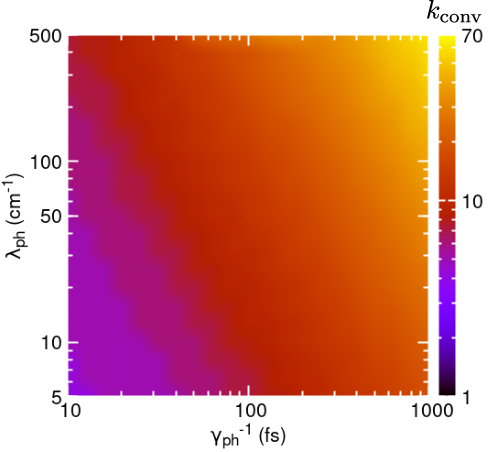}
    \caption{Heat map of the number of steps $k_\mathrm{conv}$ needed to achieve the convergence of the SCBA [$\delta\Sigma^{(k)}\leq\varepsilon_\mathrm{tol}$ for $k\geq k_\mathrm{conv}$] for different values of the reorganization energy $\lambda_\mathrm{ph}$ and the environment-reorganization timescale $\gamma_\mathrm{ph}^{-1}$.}
    \label{Fig:fig_steps_to_convergence-281024}
\end{figure}

Figure~\ref{Fig:step_vs_error_090824}(a) suggests that the convergence of the SCBA slows down with increasing the reorganization energy or the environment-reorganization timescale.
This is corroborated in Fig.~\ref{Fig:fig_steps_to_convergence-281024}, which shows the heat map of the number of iterations $k_\mathrm{conv}$ the SCBA needs to converge [$\delta\Sigma^{(k)}\leq\varepsilon_\mathrm{tol}$ for $k\geq k_\mathrm{conv}$] for different values of $\lambda_\mathrm{ph}$ and $\gamma_\mathrm{ph}^{-1}$.
Even in the regime of strong interaction ($\lambda_\mathrm{ph}=500\:\mathrm{cm}^{-1}$) with slow environment ($\gamma_\mathrm{ph}^{-1}=1\:\mathrm{ps}$), the convergence is achieved in around 70 iterations.
Interestingly, we find that $k_\mathrm{conv}$ remains essentially unaffected by variations in the excitonic parameters $J$ and $\Delta\varepsilon$ for fixed $\lambda_\mathrm{ph}$ and $\gamma_\mathrm{ph}$.
For all parameter combinations in Fig.~\ref{Fig:dimer-obo-heat-J-de}, the SCBA converges in 10--13 iterations.
The pace of the convergence appears to be relatively weakly affected by temperature variations; see Fig.~\ref{Fig:step_vs_error_090824}(a).
Figure~\ref{Fig:step_vs_error_090824}(b) suggests that this holds true even in larger aggregates interacting with structured environments [Figs.~\ref{Fig:fmo_pops}(b1) and~\ref{Fig:fmo_pops}(b2)].

\subsubsection{Overdamped phonon continuum: Self-energy superoperator}
\label{SSSec:only-phonons-self-energy}

\begin{figure*}[htbp!]
    \centering
    \includegraphics[width=\textwidth]{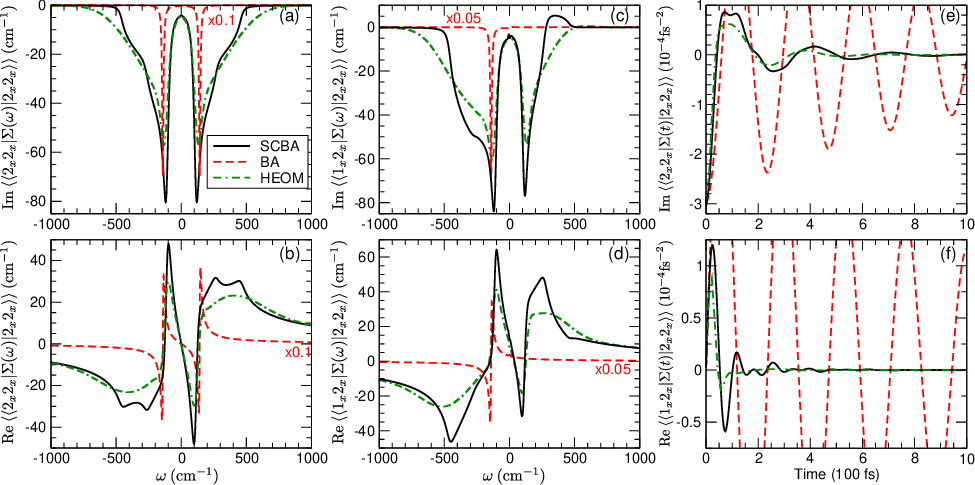}
    \caption{(a)--(d): Frequency dependence of the imaginary [(a) and (c)] and real [(b) and (d)] parts of the matrix elements of the self-energy superoperator $\Sigma(\omega)$ computed using SCBA (solid lines), BA (dashed lines), and HEOM (double dashed-dotted lines).
    We show matrix elements describing the population flux out of the higher-energy exciton state $|2_x\rangle$ [(a) and (b)] and the population-to-coherence transfer from the higher-energy exciton state [(c) and (d)].
    For visual clarity, the BA results are multiplied by a factor of (a) and (b) 0.1, and (c) and (d) 0.05.
    In panels (a)--(d), we use $\eta=1\:\mathrm{cm}^{-1}$.
    (e) and (f): Time dependence of the matrix elements of the self-energy superoperator $\Sigma(t)$ computed using SCBA, BA, and HEOM.
    The time-domain results on the timescales shown in (e) and (f) do not depend on $\eta$, see Sec.~\ref{SSec:technical}.
    Model parameters assume the same values as in Figs.~\ref{Fig:Fig_scba_ph}(b1) and~\ref{Fig:Fig_scba_ph}(b2).
    }
    \label{Fig:Fig_wt_obo}
\end{figure*}

We formulate our approximate approaches in the frequency domain, with the self-energy (memory-kernel) superoperator as their central quantity, and Figs.~\ref{Fig:Fig_wt_obo}(a)--\ref{Fig:Fig_wt_obo}(d) discuss the reflections of the above-summarized time-domain observations on the frequency domain.
We choose the slow-environment regime analyzed in Fig.~\ref{Fig:Fig_scba_ph}(b) and concentrate on the matrix elements $\llangle 2_x2_x|\Sigma(\omega)|2_x2_x\rrangle$ and $\llangle 1_x2_x|\Sigma(\omega)|2_x2_x\rrangle$ describing, respectively, the population flux out of the higher-energy exciton state and the population-to-coherence transfer from that state.
The results in Figs.~\ref{Fig:Fig_wt_obo}(a)--\ref{Fig:Fig_wt_obo}(d) are obtained setting the artificial-broadening parameter to $\eta=1\:\mathrm{cm}^{-1}$, and we have checked that varying $\eta$ over the range $[0.5,5]\:\mathrm{cm}^{-1}$ does not qualitatively (and to a large extent quantitatively) affect the results presented here. 
Within the BA, the imaginary parts of both self-energy matrix elements display very narrow peaks centered around the exciton energy gap (at $\omega=\pm \Delta\varepsilon_X$), see Figs.~\ref{Fig:Fig_wt_obo}(a) and~\ref{Fig:Fig_wt_obo}(c), which is compatible with the oscillatory features of the BA in Fig.~\ref{Fig:Fig_scba_ph}(b).
Meanwhile, the peaks of the numerically exact profiles are much wider, and their centers are somewhat shifted from $\Delta\varepsilon_X$.
While the SCBA profiles overall reasonably reproduce the numerically exact ones in terms of both peak positions and shapes, we note the tendency of the SCBA towards somewhat excessive peak shifting and narrowing.
The relative advantage of the SCBA over the BA is the most obvious for the population-to-coherence transfer, when the BA completely misses the peak appearing somewhat below $\Delta\varepsilon_X$ in both HEOM and SCBA profiles; see Fig.~\ref{Fig:Fig_wt_obo}(c).

In Figs.~\ref{Fig:Fig_wt_obo}(e) and~\ref{Fig:Fig_wt_obo}(f), we compare the BA and SCBA results for the self-energy superoperator in the time domain to the corresponding numerically exact result.
As discussed in Sec.~\ref{SSec:technical}, these time-domain results do not depend on $\eta$.
The general tendencies observed in the exciton dynamics in Fig.~\ref{Fig:Fig_scba_ph}(b) are also seen on the level of the time-dependent memory kernel.
Namely, the BA result for the time-dependent memory kernel displays pronounced weakly damped oscillatory features, while the SCBA reproduces the exact result reasonably well.
The quantitative agreement between the SCBA and HEOM results for the matrix element connected to population-to-population transfer is somewhat better than in the case of population-to-coherence transfer; compare Fig.~\ref{Fig:Fig_wt_obo}(e) to Fig.~\ref{Fig:Fig_wt_obo}(f).  
We finally note that obtaining the memory-kernel superoperator in the time domain is generally difficult.~\cite{JChemPhys.138.174103}
On the contrary, our time-domain results are readily obtained by the numerical Fourier transformation of the corresponding frequency-domain self-energies.

\subsubsection{A single underdamped vibrational mode}
\label{SSSec:one-underdamped-vib}
\begin{figure}[htbp!]
    \centering
    \includegraphics[width = \columnwidth]{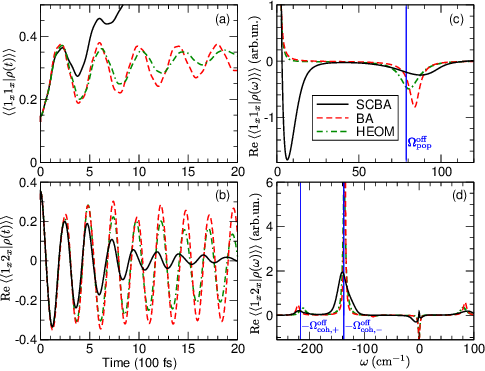}
    \caption{Time dependence of (a) the population of the lower-energy exciton state $|1_x\rangle$, and (b) the real part of the interexciton coherence computed using SCBA (solid lines), BA (dashed lines), and HEOM (double dashed-dotted lines).
    (c) and (d): The real part of the Fourier transformation of the quantities displayed in (a) and (b), respectively.
    The dimer is coupled to an underdamped vibrational mode characterized by $\omega_0=213\:\mathrm{cm}^{-1}$, $S_0=0.024$, and $\gamma_0=3\:\mathrm{cm}^{-1}$.
    The vertical lines in (c) [(d)] show the frequencies of oscillatory features in exciton-population (interexciton-coherence) dynamics obtained using the weak-interaction vibronic-exciton model, see Eqs.~\eqref{Eq:Omega_pop_off}--\eqref{Eq:Omega_coh_off_+}.
    }
    \label{Fig:Fig_omega_213_gamma_3}
\end{figure}
\begin{figure}[htbp!]
    \centering
    \includegraphics[width = \columnwidth]{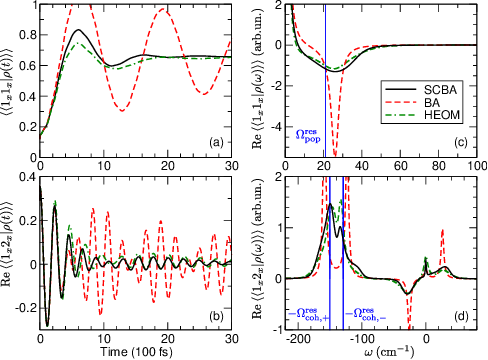}
    \caption{Time dependence of (a) the population of the lower-energy exciton state $|1_x\rangle$, and (b) the real part of the interexciton coherence computed using SCBA (solid lines), BA (dashed lines), and BA (double dashed-dotted lines).
    (c) and (d): The real part of the Fourier transformation of the quantities displayed in (a) and (b), respectively.
    The dimer is coupled to an underdamped vibrational mode characterized by $\omega_0=138\:\mathrm{cm}^{-1}$, $S_0=0.023$, and $\gamma_0=3\:\mathrm{cm}^{-1}$.
    The vertical lines in (c) [(d)] show the frequencies of oscillatory features in exciton-population (interexciton-coherence) dynamics obtained using the weak-interaction vibronic-exciton model, see Eqs.~\eqref{Eq:Omega_pop_res} and~\eqref{Eq:Omega_coh_res}.
    }
    \label{Fig:Fig_omega_138_gamma_3}
\end{figure}
Here, using the values of $J,\Delta\varepsilon,$ and $T$ summarized in Table~\ref{Tab:model_parameters}, we benchmark the (SC)BA when the SD of the exciton--environment interaction is modeled using Eq.~\eqref{Eq:def-UBO-SD}.
We set $\gamma_0=3\:\mathrm{cm}^{-1}$, corresponding to the relaxation timescale of $\gamma_0^{-1}=1.77\:\mathrm{ps}$.
Keeping in mind that the interaction with an individual intrachromophore mode is, in general, relatively weak ($S_0\sim 0.01$, irrespective of the vibrational frequency $\omega_0$),~\cite{NatCommun.13.2912} one could expect that already the BA recovers the numerically exact exciton dynamics.
Figures~\ref{Fig:Fig_omega_213_gamma_3}(a) and~\ref{Fig:Fig_omega_213_gamma_3}(b) reveal that this is indeed the case when the vibrational mode is not resonant with the excitonic energy gap ($\omega_0\neq\Delta\varepsilon_X$).
The true dynamics then exhibits weakly damped oscillations in both exciton populations and interexciton coherence, which reflect the relatively slow and inefficient interchromophore population transfer discussed in the literature.~\cite{JPhysChemLett.6.627}
While the BA reproduces the oscillatory behavior fairly well on the timescales we focus on, the SCBA produces an excessively fast oscillation damping and an overall incorrect dynamics of exciton populations.
One may then expect that the general tendency of the self-consistent cycle toward the fast equilibration of the excitonic subsystem could render the SCBA a reliable approximation when the vibrational frequency is nearly resonant with the exciton-energy gap ($\omega_0\approx\Delta\varepsilon_X$).
At resonance, the existing results show that the damping of the interexciton coherence and the concomitant interchromophore population transfer are particularly fast.~\cite{JPhysChemLett.6.627} 
Figures~\ref{Fig:Fig_omega_138_gamma_3}(a) and~\ref{Fig:Fig_omega_138_gamma_3}(b) show that the SCBA is indeed sufficiently good at reproducing this rapid equilibration of both exciton populations and coherences.

In the off-resonant case, the exciton populations and interexciton coherence predominantly oscillate at frequencies $|\omega_0-\Delta\varepsilon_X|$ and $\Delta\varepsilon_X$, respectively; see the most pronounced features of the spectra in Figs.~\ref{Fig:Fig_omega_213_gamma_3}(c) and~\ref{Fig:Fig_omega_213_gamma_3}(d).
The spectra in Figs.~\ref{Fig:Fig_omega_213_gamma_3}(c),~\ref{Fig:Fig_omega_213_gamma_3}(d),~\ref{Fig:Fig_omega_138_gamma_3}(c), and~\ref{Fig:Fig_omega_138_gamma_3}(d) originate from our frequency-domain computations so that they are broadened with the parameter $\eta$. 
The incorrect SCBA population dynamics in Fig.~\ref{Fig:Fig_omega_213_gamma_3}(a) is reflected on the frequency domain as the spurious low-frequency feature in Fig.~\ref{Fig:Fig_omega_213_gamma_3}(c).
In the resonant case, the dynamics of interexciton coherence displays beats most probably stemming from oscillations at two similar frequencies; see the most intensive features in Fig.~\ref{Fig:Fig_omega_138_gamma_3}(d).
In contrast to the off-resonant case, the relation of the beating frequency or the frequency of population oscillations in Fig.~\ref{Fig:Fig_omega_138_gamma_3}(c) to the inherent energy scales of the problem ($\Delta\varepsilon,J,\omega_0$) is not obvious.   
To establish such a relation, we note that even the weak exciton--vibration interaction can induce appreciable mixing of the vibrational and excitonic levels, and thus render the description in terms of vibronic--exciton states more appropriate.
To discuss our observations in terms of these hybrid states, we assume, for simplicity, that the vibrational mode is undamped, i.e., $\gamma_{0}=0$.
Then, the Hamiltonian is analogous to the Jaynes--Cummings Hamiltonian of quantum optics.~\cite{JChemPhys.122.134103}
When $S_0\ll 1$, it is sufficient to consider at most a single vibrational quantum,~\cite{SciRep.3.2029} and in Appendix~\ref{App:vibronic_exciton}, we conclude that the first two states above the lowest-lying vibronic-exciton state $|1_x,v_0=0\rangle$ are linear combinations of the states $|1_x,v_0=1\rangle$ and $|2_x,v_0=0\rangle$.
In the nearly resonant case ($\Delta\varepsilon_X\approx\omega_0$), the energies corresponding to the oscillatory features in exciton populations,
\begin{equation}
\label{Eq:Omega_pop_res}
    \Omega_\mathrm{pop}^\mathrm{res}\approx\omega_0\sqrt{2S_0}|\sin(2\theta)|,
\end{equation}
and the interexciton coherence,
\begin{equation}
\label{Eq:Omega_coh_res}
   \Omega_{\mathrm{coh},\pm}^\mathrm{res}\approx\Delta\varepsilon_X\pm\omega_0\sqrt{\frac{S_0}{2}}|\sin(2\theta)|
\end{equation}
are proportional to $\sqrt{S_0}$ and agree reasonably well with the corresponding numerically exact and SCBA results; see Figs.~\ref{Fig:Fig_omega_138_gamma_3}(c) and~\ref{Fig:Fig_omega_138_gamma_3}(d).
Interestingly, the frequency of the beats in the interexciton-coherence dynamics virtually coincides with the frequency of population oscillations, i.e., $\Omega_\mathrm{pop}^\mathrm{res}\approx\Omega_\mathrm{coh,+}^\mathrm{res}-\Omega_\mathrm{coh,-}^\mathrm{res}$.  
In the off-resonant case $\Delta\varepsilon_X\neq\omega_0$, we find that the frequency of population oscillations is shifted from $|\Delta\varepsilon_X-\omega_0|$ by an amount proportional to $S_0$, i.e.,
\begin{equation}
\label{Eq:Omega_pop_off}
    \Omega_\mathrm{pop}^\mathrm{off}-|\Delta\varepsilon_X-\omega_0|\approx S_0\sin^2(2\theta)\frac{\omega_0^2}{|\Delta\varepsilon_X-\omega_0|},
\end{equation}
which agrees very well with the HEOM result, and not so well with the BA and SCBA results; see Fig.~\ref{Fig:Fig_omega_213_gamma_3}(c).
The frequency shift of the most pronounced component of exciton-coherence oscillations from $\Delta\varepsilon_X$ is also linear in $S_0$,
\begin{equation}
\label{Eq:Omega_coh_off_-}
    \Omega_{\mathrm{coh},-}^\mathrm{off}-\Delta\varepsilon_X\approx-\frac{1}{2}S_0\sin^2(2\theta)\frac{\omega_0^2}{|\Delta\varepsilon_X-\omega_0|},
\end{equation}
in good agreement with the HEOM and BA results in Fig.~\ref{Fig:Fig_omega_213_gamma_3}(d).
We mention that the much less intensive feature of the coherence-oscillation spectrum appearing around the vibration energy is shifted from $\omega_0$ by 
\begin{equation}
\label{Eq:Omega_coh_off_+}
    \Omega_{\mathrm{coh},+}^\mathrm{off}-\omega_0\approx\frac{1}{2}S_0\sin^2(2\theta)\frac{\omega_0^2}{|\Delta\varepsilon_X-\omega_0|}.
\end{equation}

\subsection{Reliability of the SCBA: Seven-site model of the FMO complex}
\label{SSec:fmo}

\begin{figure*}[htbp!]
    \centering
    \includegraphics[width=0.9\textwidth]{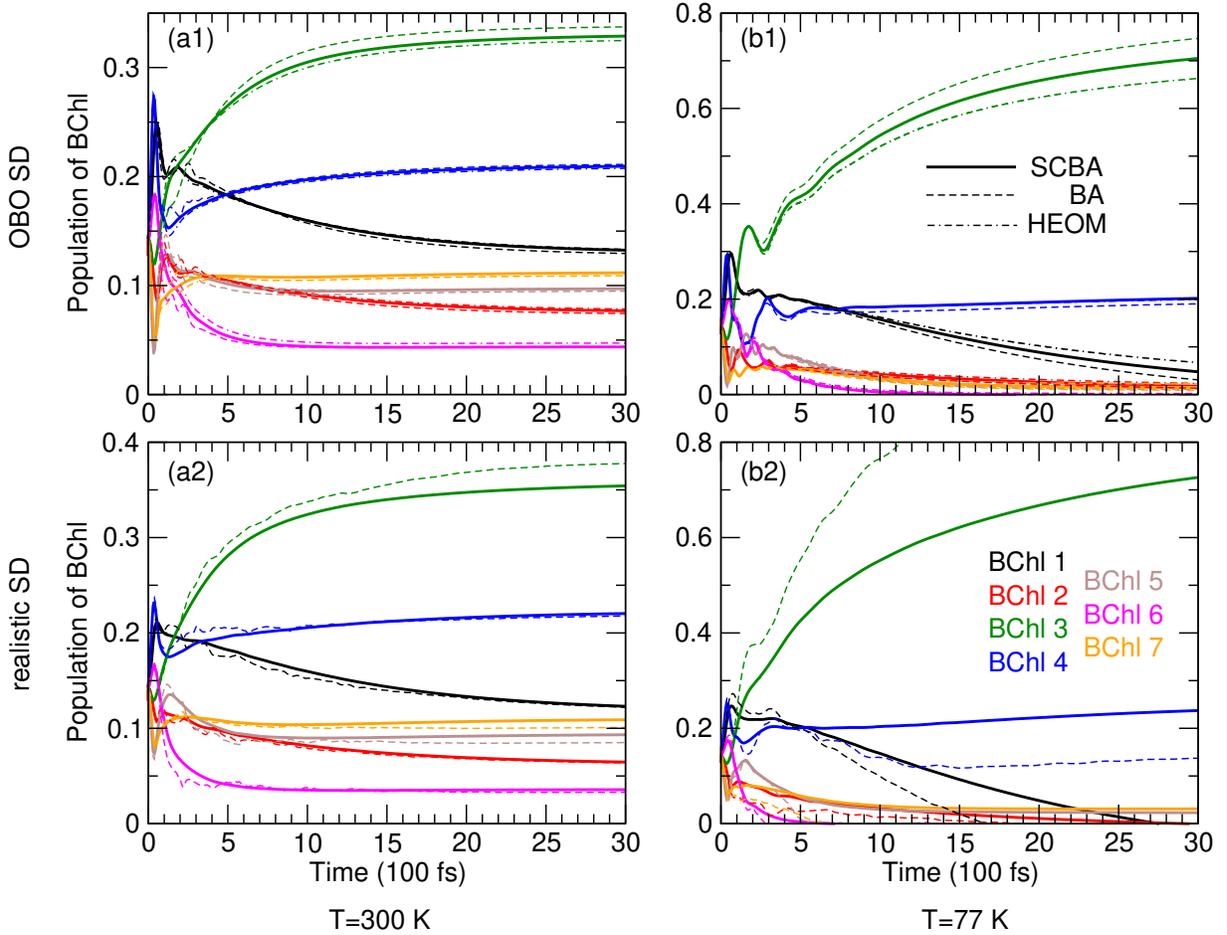}
    \caption{Comparison of SCBA (solid), BA (dashed), and HEOM (double dash-dotted) dynamics of BChl populations within the seven-site model of the FMO complex.
    Computations are performed at $T=300\:\mathrm{K}$ [(a1) and (a2)] and $T=77\:\mathrm{K}$ [(b1) and (b2)] using the OBO [(a1) and (b1)] and realistic [(a2) and (b2)] SDs of the exciton--environment interaction.
    As the numerically exact dynamics are not available for the realistic SD, panels (a2) and (b2) compare the SCBA and BA dynamics.
    }
    \label{Fig:fmo_pops}
\end{figure*}

\begin{figure*}[htbp!]
    \centering
    \includegraphics[width=0.9\textwidth]{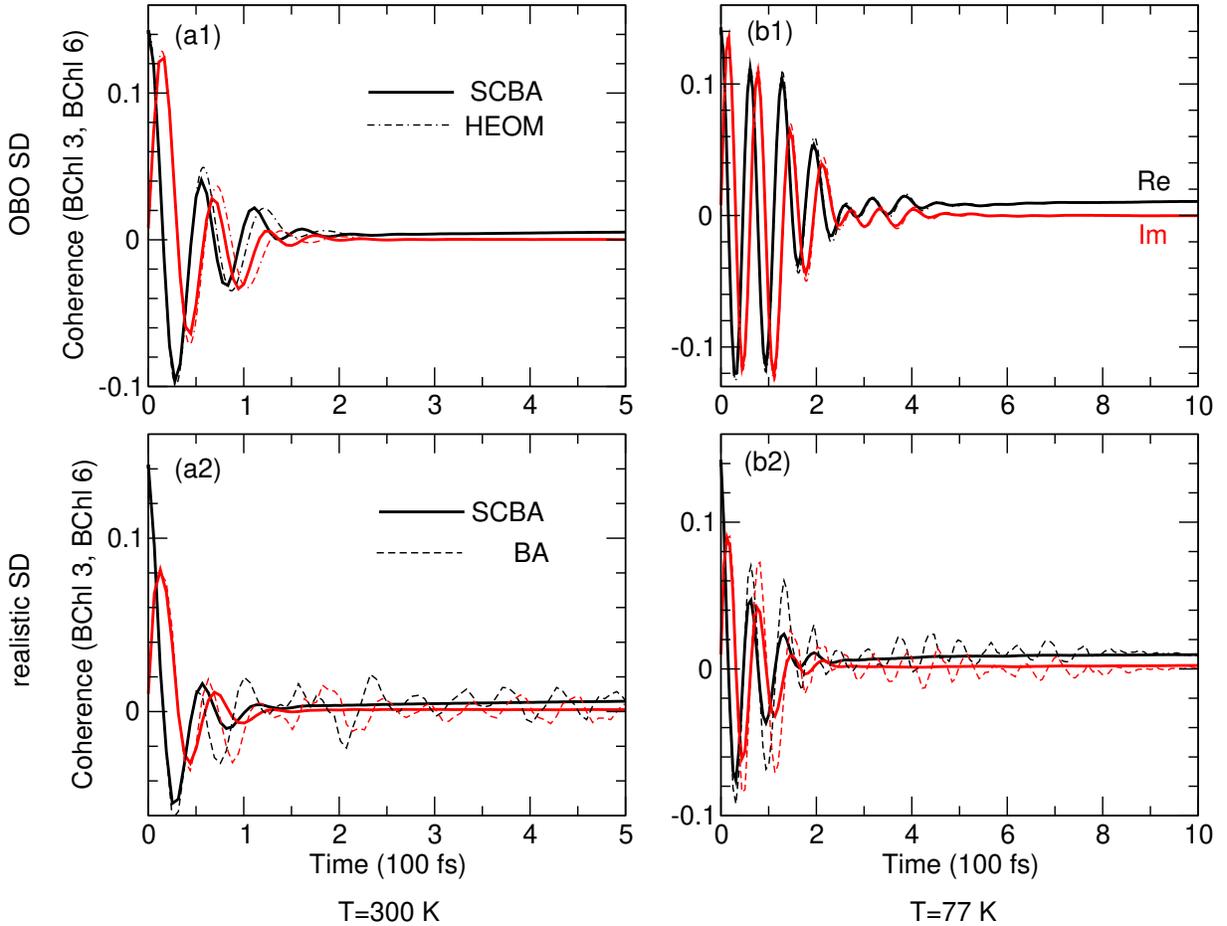}
    \caption{Comparison of SCBA (solid), BA (dashed), and HEOM (double dash-dotted) dynamics of the interchromophore coherence between BChl 3 and BChl 6 within the seven-site model of the FMO complex.
    Computations are performed at $T=300\:\mathrm{K}$ [(a1) and (a2)] and $T=77\:\mathrm{K}$ [(b1) and (b2)] using the OBO [(a1) and (b1)] and realistic [(a2) and (b2)] SDs of the exciton--environment interaction.
    In (a1) and (b1), we omit BA dynamics for visual clarity.
    As the numerically exact dynamics are not available for the realistic SD, panels (a2) and (b2) compare the SCBA and BA dynamics.}
    \label{Fig:fmo_cohs}
\end{figure*}

In Figs.~\ref{Fig:fmo_pops} and~\ref{Fig:fmo_cohs}, we benchmark the SCBA and BA on the widely studied seven-site model of the FMO complex found in green sulfur bacteria.
Detailed benchmarks of our approximations for chromophore populations (Fig.~\ref{Fig:fmo_pops}) and interchromophore coherences (Fig.~\ref{Fig:fmo_cohs}) against numerically exact results are possible for the OBO SD [Eq.~\eqref{Eq:def-OBO-SD}].
In Figs.~\ref{Fig:fmo_pops}(a1),~\ref{Fig:fmo_pops}(b1),~\ref{Fig:fmo_cohs}(a1), and~\ref{Fig:fmo_cohs}(b1), we use $\lambda_\mathrm{ph}=35\:\mathrm{cm}^{-1}$ and $\gamma_\mathrm{ph}^{-1}=50\:\mathrm{fs}$ ($\gamma_\mathrm{ph}=106.2\:\mathrm{cm}^{-1}$).
To explore the viability of our methodology, we also apply it to the model using the structured SD emerging from atomistic simulations performed in Ref.~\onlinecite{JPhysChemLett.7.3171}.
As we are not aware of any numerically exact results for the dynamics modulated by this structured bath, Figs.~\ref{Fig:fmo_pops}(a2),~\ref{Fig:fmo_pops}(b2),~\ref{Fig:fmo_cohs}(a2), and~\ref{Fig:fmo_cohs}(b2) compare the SCBA and BA results.
Details on the excitonic Hamiltonian $H_S$ and the structured SD used are summarized in Appendix~\ref{App:FMO-details}.

Using the OBO SD, the BA already provides a good approximation to the subpicosecond dynamics of chromophore populations [Figs.~\ref{Fig:fmo_pops}(a1) and~\ref{Fig:fmo_pops}(b1)], in agreement with the findings of Ref.~\onlinecite{PhysRevE.86.011915}.
Nevertheless, on a picosecond timescale, the BA overestimates (underestimates) populations of low-energy (high-energy) states, and this effect becomes more pronounced with decreasing the temperature; compare Fig.~\ref{Fig:fmo_pops}(a1) to Fig.~\ref{Fig:fmo_pops}(b1).
While the SCBA suffers from the same deficiency, its predictions for chromophore populations are systematically closer to the numerically exact results throughout the time window examined.
At higher temperatures, Figs.~\ref{Fig:fmo_pops}(a1) and~\ref{Fig:fmo_cohs}(a1) suggest that the SCBA is better at approximating population dynamics than coherence dynamics.
Meanwhile, Figs.~\ref{Fig:fmo_pops}(b1) and~\ref{Fig:fmo_cohs}(b1) suggest that the reverse is true at lower temperatures.
Figures~\ref{Fig:fmo_cohs}(a1) and~\ref{Fig:fmo_cohs}(b1) do not show BA results, which, at all temperatures, display oscillatory features lasting much longer than the numerically exact method predicts.  

Comparing panels (a2) and (a1) [(b2) and (b1)] in Figs.~\ref{Fig:fmo_pops} and~\ref{Fig:fmo_cohs}, we conclude that our approximations deliver reasonable results for exciton dynamics in the structured environment, as it is overall similar to the dynamics in the featureless environment.
The problems with longer-time population dynamics of extremal-energy states are exacerbated in the structured environment at lower temperatures, when SCBA (and also BA) predicts nonphysical (greater than 1 or negative) populations of such states; see Fig.~\ref{Fig:fmo_pops}(b2).
A possible origin of these problems can be understood from our analysis of the dynamics modulated by an underdamped vibrational mode; see Sec.~\ref{SSSec:one-underdamped-vib}.
There, we find that the resonance condition between the exciton-energy gap and the vibrational energy quantum is crucial to the success of the SCBA.
Here, exciton-energy gaps fluctuate due to low-frequency phonon modes, their fluctuations becoming more pronounced with increasing temperature.
Therefore, at higher temperatures, satisfying resonance conditions between exciton-energy gaps and weakly damped vibrational modes is more probable than at lower temperatures.
In other words, the SCBA is expected to work better at higher temperatures.
The SCBA coherence dynamics in Figs.~\ref{Fig:fmo_cohs}(a2) and~\ref{Fig:fmo_cohs}(b2) is physically sensible, while the lifetime of interchromophore coherences in the structured environment is somewhat shorter than in the featureless environment at both temperatures examined.

\section{Summary and outlook}
\label{Sec:summary}
We have developed and benchmarked the self-consistent Born approximation for studying the dynamics of EET through a multichromophoric aggregate linearly interacting with a bosonic environment.
We start from the lowest-order approximation for the memory kernel of the QME and improve it in the self-consistent cycle based on the QME represented in the Liouville space.
We find that the SCBA reproduces the exact exciton dynamics modulated by an overdamped phonon continuum very well, even in the generally difficult regimes of strong exciton--environment interaction, slow environmental reorganization, and low temperature.
This success of the SCBA can be understood from the analytically tractable example---coherence-dephasing dynamics in the pure-dephasing model.
We conclude that the reliability of the SCBA when the dynamics is modulated by an underdamped vibrational mode leans on the resonance between the mode frequency and the exciton-energy gap.
Nevertheless, in a structured environment comprising both an overdamped phonon continuum and a number of underdamped vibrational modes, the SCBA reasonably describes exciton dynamics through the seven-site model of the FMO complex.

Importantly, our method does not introduce any assumptions on the form of the exciton--environment SD, making it a strong candidate for studying exciton dynamics modulated by structured environments whose properties are extracted from experiments or atomistic simulations.
Although the method sometimes leads to unphysical results [see, for example, Fig.~\ref{Fig:fmo_pops}(b2)], it can, in principle, be improved by enlarging the set of the diagrams included in Green's superoperator beyond that in Fig.~\ref{Fig:ba_scba_final}(b2).
Namely, Fig.~\ref{Fig:fmo_pops}(b2) shows that the onset of the unphysical behavior in the SCBA result is shifted towards later times with respect to the BA result so that including additional diagrams can be expected to further improve the SCBA result. 
Systematic improvements are possible by performing the self-consistent cycle starting from a higher-order approximation for the memory kernel.
The next member of the family of self-consistent approximations thus obtained is the so-called one-crossing approximation, in which the starting memory kernel is the sum of the first two diagrams in Fig.~\ref{Fig:diags_final}(f).
The final memory kernel then contains all diagrams in which the lines representing the environmental assistance cross at most once.
However, the one-crossing approximation is computationally much more demanding than the SCBA, as each iteration involves a double integral over frequency.
The practical applicability of the one-crossing approximation is thus determined by the balance between its computational requirements and the improvements it offers over the SCBA.

Remarkably, it may happen that the good results obtained in the lowest-order Born approximation are deteriorated by including the higher-order diagrams captured by the SCBA; see, for example, Fig.~\ref{Fig:Fig_omega_213_gamma_3}(a). %(molecular dimer interacting with \emph{a single underdamped} mode).
Such situations typically feature underdamped (or undamped) environments, whose SDs are narrow (the corresponding bath correlation functions decay very slowly).
Then, we have to recall that the perturbation series in Fig.~\ref{Fig:diags_final}(d) is actually an asymptotic series, which in general does not converge.
Nevertheless, in realistic environments, overdamped contributions may be considered to come to rescue the convergence of the perturbation series, as these effectively hide the unphysical features caused by underdamped contributions alone.

Finally, our Liouville-space frequency-domain formulation of the SCBA suggests that it might be used as a computationally efficient and reasonably accurate approach to compute experimentally accessible nonlinear response functions.
To test such a possibility, one should generalize the present formulation so that memory kernels in different excited-state sectors can be computed. 

\section*{Author Contributions}
V.J. developed the method and performed all analytical derivations.
V.J. and T.M. numerically implemented the method and conceived the examples on which the method was tested.
V.J. performed all numerical computations and prepared the initial version of the manuscript.
Both authors contributed to the final version of the manuscript.

\acknowledgments
The work in Prague is funded by the Czech Science Foundation (GA\v CR), Grant No. 22-26376S.
The work in Belgrade is supported by the Institute of Physics Belgrade, through the grant by the Ministry of Science, Technological Development, and Innovation of the Republic of Serbia.
Numerical computations were performed on the PARADOX-IV supercomputing facility in the Scientific Computing Laboratory, National Center of Excellence for the Study of Complex Systems, Institute of Physics Belgrade.

\section*{Data availability}
The data that support the findings of this study are available from the corresponding author upon reasonable request.

\begin{widetext}
\appendix
\section{Coherence dephasing in the pure-dephasing model (OBO SD and high-temperature limit)}\label{App:analytics}
We first show that the CFE of the exact $\Sigma(\omega)$ is given by Eq.~\eqref{Eq:exact_CFE}.
In the high-temperature limit, the real part of the lineshape function can be approximated as~\cite{Mukamel-book}
\begin{equation}
\label{Eq:lineshape_high_T}
    g^r(t)=\frac{2\lambda_\mathrm{ph}T}{\gamma_\mathrm{ph}^2}(e^{-\gamma_\mathrm{ph}t}+\gamma_\mathrm{ph}t-1).
\end{equation}
Inserting Eq.~\eqref{Eq:lineshape_high_T} into Eq.~\eqref{Eq:exact_G_t_pure_phase_noise} and performing the Fourier transformation of the latter gives ($n_2\neq n_1$)
\begin{equation}
\label{Eq:cal_G_w_before_CFE}
\llangle n_2n_1|\mathcal{G}(\omega)|n_2n_1\rrangle=e^{\Lambda_\mathrm{ph}}\sum_{k=0}^{+\infty}\frac{(-\Lambda_\mathrm{ph})^k}{k!}\frac{1}{\omega-\varepsilon_{n_2n_1}+i\Lambda_\mathrm{ph}\gamma_\mathrm{ph}+ik\gamma_\mathrm{ph}},
\end{equation}
where we introduce the dimensionless parameter $\Lambda_\mathrm{ph}=\frac{4\lambda_\mathrm{ph}T}{\gamma_\mathrm{ph}^2}$.
It is known that the zero-temperature absorption lineshape (the excitation-addition spectral function) of a two-level system whose energy gap $\varepsilon_{eg}$ between the ground and excited states is modulated by an undamped vibrational mode (frequency $\omega_0$ and HR factor $S_0$) is proportional to the negative imaginary part of the retarded Green's function (see, for example, Chap.~8 of Ref.~\onlinecite{Mukamel-book} or Chap.~4 of Ref.~\onlinecite{Mahanbook}),
\begin{equation}
\label{Eq:G_R_before_CFE}
    G^R(\omega)=e^{-S_0}\sum_{k=0}^{+\infty}\frac{S_0^{k}}{k!}\frac{1}{\omega-\varepsilon_{eg}+S_0\omega_0-k\omega_0+i\eta}.
\end{equation}
The CFE of Eq.~\eqref{Eq:G_R_before_CFE} reads as~\cite{PhysRevB.56.4494,PhysRevB.74.245104}
\begin{equation}
\label{Eq:G_R_CFE}
    G^R(\omega)=\cfrac{1}{\omega-\varepsilon_{eg}-\cfrac{S_0\omega_0^2}{\omega-\varepsilon_{eg}-\omega_0-\cfrac{2S_0\omega_0^2}{\omega-\varepsilon_{eg}-2\omega_0-\cdots}}}.
\end{equation}
The CFE of $\llangle n_2n_1|\mathcal{G}(\omega)|n_2n_1\rrangle$ is then obtained by substituting $\omega-\varepsilon_{eg}\to\omega-\varepsilon_{n_2n_1}$, $S_0\to-\Lambda_\mathrm{ph}$, and $\omega_0\to-i\gamma_\mathrm{ph}$ in Eq.~\eqref{Eq:G_R_CFE}, which follows from comparing Eq.~\eqref{Eq:cal_G_w_before_CFE} with Eq.~\eqref{Eq:G_R_before_CFE}, and reads
\begin{equation}
\label{Eq:G_R_CFE_here}
    \llangle n_2n_1|\mathcal{G}(\omega)|n_2n_1\rrangle=\cfrac{1}{\omega-\varepsilon_{n_2n_1}-\cfrac{4\lambda_\mathrm{ph}T}{\omega-\varepsilon_{n_2n_1}+i\gamma_\mathrm{ph}-\cfrac{2\cdot 4\lambda_\mathrm{ph}T}{\omega-\varepsilon_{n_2n_1}+2i\gamma_\mathrm{ph}-\cdots}}}.
\end{equation}
Keeping in mind that $\llangle n_2n_1|\Sigma(\omega)|n_2n_1\rrangle=\omega-\varepsilon_{n_2n_1}-\llangle n_2n_1|\mathcal{G}(\omega)|n_2n_1\rrangle^{-1}$, see Eq.~\eqref{Eq:Dyson-freq}, one immediately obtains Eq.~\eqref{Eq:exact_CFE}.

Working in the high-temperature limit, Tanimura and Kubo obtained an expression very similar to Eq.~\eqref{Eq:G_R_CFE_here}; see Appendix~B of Ref.~\onlinecite{JPhysSocJpn.58.101}.
They considered a single two-level chromophore (labeled $n_1$) and its optical coherence, whose time evolution is governed by $e^{-i\varepsilon_{n_1}t}e^{-g(t)}$.~\cite{Mancal_2014}
The numerators of their CFE feature nonzero imaginary parts originating from $g^i(t)$.
Here, however, we consider the interchromophore coherence between excitonically uncoupled chromophores $n_2$ and $n_1$ and assume that their environments are identical and uncorrelated.
In the time domain, the coherence involves the de-excitation of chromophore $n_1$ and excitation of chromophore $n_2$, so that it evolves according to $e^{-i\varepsilon_{n_2}t}e^{-g(t)}[e^{-i\varepsilon_{n_1}t}e^{-g(t)}]^*=e^{-i\varepsilon_{n_2n_1}t}e^{-2g^r(t)}$.
The imaginary parts of the line shape functions cancel out, in agreement with Eq.~\eqref{Eq:exact_G_t_pure_phase_noise}, rendering the numerators of the CFE in Eqs.~\eqref{Eq:G_R_CFE_here} and~\eqref{Eq:exact_CFE} purely real.

We now argue that the CFE of the SCBA self-energy is given by Eq.~\eqref{Eq:SCBA_CFE}.
Our starting point is the BA propagator in the frequency domain ($n_2\neq n_1$),
\begin{equation}
\label{Eq:BA_propagator_w}
    \llangle n_2n_1|\mathcal{G}_\mathrm{BA}(\omega)|n_2n_1\rrangle=\cfrac{1}{\omega-\varepsilon_{n_2n_1}-\cfrac{4\lambda_\mathrm{ph}T}{\omega-\varepsilon_{n_2n_1}+i\gamma_\mathrm{ph}}},
\end{equation}
which is obtained by inserting the BA self-energy [Eq.~\eqref{Eq:BA_sigma_pure_phase_noise}] into the Dyson equation [Eq.~\eqref{Eq:Dyson-freq}].
The BA propagator is then inserted into Eq.~\eqref{Eq:SCBA-self-energy-freq} to obtain the following self-energy $\Sigma^{(2)}$ in the next iteration:
\begin{equation}
    \llangle n_2n_1|\Sigma^{(2)}(\omega)|n_2n_1\rrangle=\int_{-\infty}^{+\infty}\frac{d\nu}{2\pi}\frac{4\lambda_\mathrm{ph}\gamma_\mathrm{ph}T}{(\omega-\nu)^2+\gamma_\mathrm{ph}^2}\llangle n_2n_1|\mathcal{G}_\mathrm{BA}(\omega)|n_2n_1\rrangle.
\end{equation}
Here, we have approximated $\mathrm{coth}\left(\frac{\beta(\omega-\nu)}{2}\right)\approx\frac{2T}{\omega-\nu}$.
The last integral is solved by integrating along a contour that is closed in the upper half-plane to ensure causality.
As the poles of the BA propagator are in the lower half-plane by construction, the only pole of the integrand in the upper half-plane is at $\nu=\omega+i\gamma_\mathrm{ph}$.
Evaluating the corresponding residue, we obtain
\begin{equation}
    \llangle n_2n_1|\Sigma^{(2)}(\omega)|n_2n_1\rrangle=\cfrac{4\lambda_\mathrm{ph}T}{\omega-\varepsilon_{n_2n_1}+i\gamma_\mathrm{ph}-\cfrac{4\lambda_\mathrm{ph}T}{\omega-\varepsilon_{n_2n_1}+2i\gamma_\mathrm{ph}}}.
\end{equation}
Repeating the above-described procedure \emph{ad infinitum}, we obtain Eq.~\eqref{Eq:SCBA_CFE}.

We end this section by presenting analytical results for the matrix elements of Green's superoperator that determine the dynamics of coherence dephasing within the BA and the Redfield theory; see Eqs.~\eqref{Eq:coh-deph-BA} and~\eqref{Eq:coh-deph-Redfield}.
Equation~\eqref{Eq:BA_propagator_w} implies that the BA propagator in the time domain can be expressed as
\begin{equation}
\label{Eq:BA_propagator_t_before_int}
    \llangle n_2n_1|\mathcal{G}_\mathrm{BA}(t)|n_2n_1\rrangle=e^{-i\varepsilon_{n_2n_1}t}\int_{-\infty}^{+\infty}\frac{d\Omega}{2\pi}e^{-i\Omega t}\frac{\Omega+i\gamma_\mathrm{ph}}{(\Omega-\Omega_+)(\Omega-\Omega_-)},
\end{equation}
where
\begin{equation}
    \Omega_{\pm}=\frac{\gamma_\mathrm{ph}}{2}\left[\pm\sqrt{\frac{16\lambda_\mathrm{ph}T}{\gamma_\mathrm{ph}^2}-1}-i\right]\approx\pm 2\sqrt{\lambda_\mathrm{ph}T}-i\frac{\gamma_\mathrm{ph}}{2}
\end{equation}
are the roots of the quadratic equation $\Omega^2+i\gamma_\mathrm{ph}\Omega-4\lambda_\mathrm{ph}T=0$, of which both lie in the lower half-plane.
We use $2\pi T/\gamma_\mathrm{ph}\gg 1$ to obtain $\Omega_\pm$ in the high-temperature limit.
The integral in Eq.~\eqref{Eq:BA_propagator_t_before_int} is solved using the contour integration with the final result,
\begin{equation}
\label{Eq:BA-exact}
 \llangle n_2n_1|\mathcal{G}_\mathrm{BA}(t)|n_2n_1\rrangle=-i\theta(t)\left[\frac{\Omega_++i\gamma_\mathrm{ph}}{\Omega_+-\Omega_-}e^{-i\Omega_+t}-\frac{\Omega_-+i\gamma_\mathrm{ph}}{\Omega_+-\Omega_-}e^{-i\Omega_-t}\right].   
\end{equation}
Retaining the contributions that are the most dominant in the high-temperature limit $2\pi T/\gamma_\mathrm{ph}\gg 1$, Eq.~\eqref{Eq:BA-exact} can be further simplified to
\begin{equation}
\begin{split}
  \llangle n_2n_1|\mathcal{G}_\mathrm{BA}(t)|n_2n_1\rrangle=-i\theta(t)e^{-i\varepsilon_{n_2n_1}t}\cos\left(2\sqrt{\lambda_\mathrm{ph}T}\:t\right)e^{-\gamma_\mathrm{ph}t/2}.  
\end{split}
\end{equation}
Equation~\eqref{Eq:Sigma_Red_from_BA} implies that the Redfield propagator in the time domain can be expressed as
\begin{equation}
\begin{split}
    \llangle n_2n_1|\mathcal{G}_\mathrm{Red}(t)|n_2n_1\rrangle&=e^{-i\varepsilon_{n_2n_1}t}\int_{-\infty}^{+\infty}\frac{d\Omega}{2\pi}\frac{e^{-i\Omega t}}{\Omega+i\frac{4\lambda_\mathrm{ph}T}{\gamma_\mathrm{ph}}}\\
    &=-i\theta(t)e^{-i\varepsilon_{n_2n_1}t}\exp\left(-\frac{4\lambda_\mathrm{ph}T}{\gamma_\mathrm{ph}}t\right).
\end{split}
\end{equation}
The approximation to the exact coherence-dephasing dynamics embodied in Eq.~\eqref{Eq:coh-deph-exact} is obtained by inserting the short-time approximation $g^r(t)\approx \lambda_\mathrm{ph}Tt^2$ to the line shape function [Eq.~\eqref{Eq:lineshape_high_T}] into Eq.~\eqref{Eq:exact_G_t_pure_phase_noise}.

\section{Vibronic--exciton model}
\label{App:vibronic_exciton}
We consider the model dimer, Eq.~\eqref{Eq:def-H_S-dimer}, in which the excitons interact with an undamped vibrational mode of frequency $\omega_0$ and HR factor $S_0$ [we set $\gamma_0=0$ in Eq.~\eqref{Eq:def-UBO-SD}].
We assume that $S_0\ll 1$ so that the reorganization energy $S_0\omega_0$ is much smaller than all the other energy scales in the problem ($\Delta\varepsilon,J$).
It is known that the center-of-mass motion of the intrachromophore vibrations, described by $B_+=(b_1+b_2)/\sqrt{2}$, does not affect the single-exciton dynamics.~\cite{NatCommun.13.2912,ProcNatlAcadSci.110.1203}
The energies of the exciton states, as well as transitions between them, are then modulated by the relative motion of intrachromophore vibrations, which is described by $B_-=(b_1-b_2)/\sqrt{2}$.
The single-exciton dynamics is governed by the Hamiltonian $H=H_X+H_{B_-}+H_{X-B_-}$, where
\begin{equation}
    H_X=\sum_{k=1}^2|k_x\rangle\left[\varepsilon_{k_x}+(-1)^k\cos(2\theta)\omega_0\sqrt{\frac{S_0}{2}}(B_-^\dagger+B_-)\right]\langle k_x|,
\end{equation}
\begin{equation}
    H_{B_-}=\omega_0 B_-^\dagger B_-,
\end{equation}
\begin{equation}
    H_{X-B_-}=-\sin(2\theta)\omega_0\sqrt{\frac{S_0}{2}}\left(|2_x\rangle\langle 1_x|+|1_x\rangle\langle 2_x|\right)(B_-^\dagger+B_-).
\end{equation}
The energy of the bare exciton state $k=1,2$ is $\varepsilon_{k_x}=\frac{\Delta\varepsilon}{2}\left[1+(-1)^k\sqrt{1+\tan^2(2\theta)}\right]$.
The dynamic modulation of the exciton energy can be taken into account exactly by transferring to the polaron frame, $\widetilde{H}=UHU^\dagger$, using the polaron transformation,
\begin{equation}
    U=\sum_{k=1}^2|k_x\rangle\langle k_x|e^{(-1)^kS_{\theta}},\quad S_{\theta}=\cos(2\theta)\sqrt{\frac{S_0}{2}}(B_-^\dagger-B_-).
\end{equation}
The bare exciton energies $\varepsilon_{k_x}$ are then shifted to $\widetilde{\varepsilon}_{k_x}=\varepsilon_{k_x}-\frac{S_0\omega_0}{2}\cos^2(2\theta)$, while the eigenstates $|k_x,v_0\rangle$ of the transformed Hamiltonian $\widetilde{H}_X+\widetilde{H}_{B_-}$ are enumerated by the exciton number $k$ and the number $v_0=0,1,\dots$ of excited vibrational quanta.
In more detail,
\begin{equation}
    |k_x,v_0\rangle=|k_x\rangle\frac{(B_-^\dagger)^{v_0}}{\sqrt{v_0!}}|\emptyset\rangle,\quad  \widetilde{\varepsilon}_{k,v_0}=\varepsilon_{k_x}-\frac{S_0\omega_0}{2}\cos^2(2\theta)+v_0\omega_0.
\end{equation}
As we assume that $S_0\ll 1$, the leading contribution (proportional to $\sqrt{S_0}$) to the transformed interaction term $\widetilde{H}_{X-B_-}$ in powers of $S_0$ is identical to $H_{X-B_-}$.
To proceed further, we additionally perform the rotating-wave approximation, which amounts to
\begin{equation}
    \widetilde{H}_{X-B_-}\approx -\omega_0\sqrt{\frac{S_0}{2}}\left(|2_x\rangle\langle 1_x| B_-+|1_x\rangle\langle 2_x|B_-^\dagger\right).
\end{equation}
We now limit ourselves to the subspace containing at most a single vibrational excitation, where we find that $\widetilde{H}_{X-B_-}$ mixes the states $|1_x,v_0=1\rangle$ and $|2_x,v_0=0\rangle$.
In the resonant case $\omega_0\approx\Delta\varepsilon_X$, these two states are nearly degenerate, and $\widetilde{H}_{X-B_-}$ lifts this degeneracy.
While there is no degeneracy in the off-resonant case $\omega_0\neq\Delta\varepsilon_X$, the vibronic mixing still affects the energy difference between these two states.
The energies of vibronically mixed states measured with respect to the energy of the lowest-lying state $|1_x,v_0=0\rangle$ then read as
\begin{equation}
    E_{v_0=1,\pm}-E_{v_0=0}=\frac{1}{2}\left[\Delta\varepsilon_X+\omega_0\pm\sqrt{(\Delta\varepsilon_X-\omega_0)^2+2S_0\omega_0^2\sin^2(2\theta)}\right].
\end{equation}
The exciton populations then exhibit oscillatory features of frequency
\begin{equation}
    \Omega_\mathrm{pop}=E_{v_0=1,+}-E_{v_0=1,-}=\sqrt{(\Delta\varepsilon_X-\omega_0)^2+2S_0\omega_0^2\sin^2(2\theta)},
\end{equation}
while the oscillations in interexciton coherence have the frequencies
\begin{equation}
 \Omega_{\mathrm{coh},\pm}=E_{v_0=1,\pm}-E_{v_0=0}.   
\end{equation}

In the nearly resonant case ($\Delta\varepsilon_X-\omega_0\approx 0$), the frequencies of oscillatory features in exciton-population and interexciton-coherence dynamics are given by Eqs.~\eqref{Eq:Omega_pop_res} and~\ref{Eq:Omega_coh_res}, respectively. 
To be consistent with the simplifications made up to now, our considerations in the off-resonant case have to keep only the lowest-order term in small $S_0$ so that
\begin{equation}
\label{Eq:pop_off_appendix}
    \Omega_\mathrm{pop}^\mathrm{off}\approx|\Delta\varepsilon_X-\omega_0|\left[1+S_0\sin^2(2\theta)\frac{\omega_0^2}{(\Delta\varepsilon_X-\omega_0)^2}\right],
\end{equation}
\begin{equation}
\label{Eq:coh_off_appendix}
    \Omega_{\mathrm{coh},\pm}^\mathrm{off}=\frac{1}{2}\left[\Delta\varepsilon_X+\omega_0\pm|\Delta\varepsilon_X-\omega_0|\pm S_0\sin^2(2\theta)\frac{\omega_0^2}{|\Delta\varepsilon_X-\omega_0|}\right].
\end{equation}
Equations~\eqref{Eq:Omega_pop_off},~\eqref{Eq:Omega_coh_off_-}, and~\eqref{Eq:Omega_coh_off_+} are then readily obtained from Eqs.~\eqref{Eq:pop_off_appendix} and~\eqref{Eq:coh_off_appendix}.
\section{Computations on the seven-site model of the FMO complex}
\label{App:FMO-details}
The excitonic Hamiltonian $H_S$ is taken from Ref.~\onlinecite{BiophysJ.91.2778}, and the values of interchromophore couplings and average chromophore energies are summarized in Table~\ref{Tab:FMO}.
\begin{table}[htbp!]
    \centering
    \begin{tabular}{c c c c c c c c}
        BChl & 1 & 2 & 3 & 4 & 5 & 6 & 7\\
        \hline
        1 & 200 & $-87.7$ & 5.5 & $-5.9$ & 6.7 & $-13.7$ & $-9.9$\\
        2 &  & 320 & 30.8 & 8.2 & 0.7 & 11.8 & 4.3\\
        3 &  &  & 0 & $-53.5$ & $-2.2$ & $-9.6$ & 6.0\\
        4 &  &  &  & 110 & $-70.7$ & $-17.0$ & $-63.3$\\
        5 &  &  &  &  & 270 & 81.1 & $-1.3$\\
        6 &  &  &  &  &  & 420 & 39.7\\
        7 &  &  &  &  &  &  & 230
    \end{tabular}
    \caption{Interchromophore couplings and average chromophore energies (in $\mathrm{cm}^{-1}$) within the seven-site model of the FMO complex considered in Ref.~\onlinecite{BiophysJ.91.2778}.
    We subtract the energy of the lowest-lying BChl3 ($12210\:\mathrm{cm}^{-1}$) from the diagonal entries.}
    \label{Tab:FMO}
\end{table}

We take the structured SD of the exciton--environment interaction from the spreadsheet \texttt{jz6b01440\_si\_002.xlsx} appearing in the Supporting Information to Ref.~\onlinecite{JPhysChemLett.7.3171}.
More specifically, we use the data from the sheet \texttt{PBE0-FMO\_subunitA-J}, which reports the total SD (comprising both the interchromophore and intrachromophore contributions) for individual BChls in one of the FMO subunits.
In our computations, we assume that the SDs for all BChls are identical and thus use the arithmetic average of the SD data for BChl1,\dots,BChl7.
The SD thus obtained is plotted in Fig.~\ref{Fig:realistic_sd}.
The frequency grid is equidistant, with spacing $\Delta\omega=0.53\:\mathrm{cm}^{-1}$, which we find sufficiently fine for the numerical integration of Eq.~\eqref{Eq:SCBA-self-energy-freq} using the ordinary trapezoidal rule.
The SD is Ohmic, and we obtain its linear behavior around $\omega=0$ by fitting the points $(0,0),(\Delta\omega,\mathcal{J}(\Delta\omega)),\dots,(5\Delta\omega,\mathcal{J}(5\Delta\omega))$ to a straight line.
As the SD is available up to $\omega_\mathrm{avail}=2650.8\:\mathrm{cm}^{-1}$, our numerical computations use $\omega_\mathrm{max}=2\omega_\mathrm{avail}$ and assume $\mathcal{J}(\omega)=0$ for $\omega_\mathrm{avail}<\omega<\omega_\mathrm{max}$.

\begin{figure}[htbp!]
    \centering
    \includegraphics[width=0.4\textwidth]{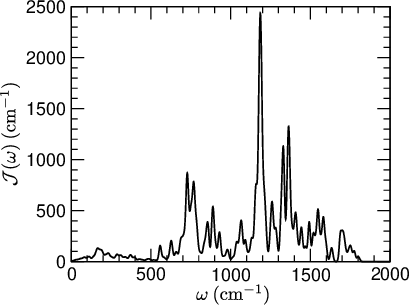}
    \caption{Spectral density of the exciton--environment interaction used to obtain the results presented in Figs.~\ref{Fig:fmo_pops} and~\ref{Fig:fmo_cohs}. The SD is extracted from the data accompanying Ref.~\onlinecite{JPhysChemLett.7.3171}.}
    \label{Fig:realistic_sd}
\end{figure}

While we take the excitonic and the exciton--environment interaction parameters from different studies and different (seven-site~\cite{BiophysJ.91.2778} and eight-site~\cite{JPhysChemLett.7.3171}) FMO models, we emphasize that our main goal is to examine the applicability of our approximate methods to a relatively realistic model of a multichromophoric complex.
The results in Figs.~\ref{Fig:fmo_pops} and~\ref{Fig:fmo_cohs} indeed suggest that our approximate methods can provide decent results on exciton dynamics in realistic models.   
\end{widetext}
\bibliography{aipsamp}% Produces the bibliography via BibTeX.
\end{document}